\documentclass[ajl]{emulateapj}

\usepackage{graphicx,epsfig,natbib,color}
\usepackage{lscape}
\usepackage{graphicx}
\usepackage{epsfig}
\usepackage{natbib}
\usepackage[colorlinks, citecolor=blue]{hyperref}

\slugcomment{}

\begin{document}
\title{Observations of Turbulent Magnetic Reconnection Within a Solar Current Sheet}

\author{X. Cheng$^{1}$, Y. Li$^{2}$, L. F. Wan$^{1}$, M. D. Ding$^{1}$, P. F. Chen$^{1}$, J. Zhang$^{3}$ \& J. J. Liu$^{4}$\\}

\affil{$^1$School of Astronomy and Space Science, Nanjing University, Nanjing 210023, China\\}\email{xincheng@nju.edu.cn}

\affil{$^2$Key Laboratory of Dark Matter and Space Astronomy, Purple Mountain Observatory, Nanjing 210008, China\\}

\affil{$^3$Department of Physics and Astronomy, George Mason University, Fairfax, VA 22030, USA\\}

\affil{$^4$Solar Physics and Space Plasma Research Center, School of Mathematics and Statistics, University of Sheffield, Sheffield S3 7RH, UK\\}

\begin{abstract}
Magnetic reconnection is a fundamental physical process in various astrophysical, space, and laboratory environments. Many pieces of evidence for magnetic reconnection have been uncovered. However, its specific processes that could be fragmented and turbulent have been short of direct observational evidence. Here, we present observations of a super-hot current sheet during SOL2017-09-10T X8.2-class solar flare that display the fragmented and turbulent nature of magnetic reconnection. As bilateral plasmas converge toward the current sheet, significant plasma heating and non-thermal motions are detected therein. Two oppositely directed outflow jets are intermittently expelled out of the fragmenting current sheet, whose intensity shows a power-law distribution in spatial frequency domain. The intensity and velocity of the sunward outflow jets also display a power-law distribution in temporal frequency domain. The length-to-width ratio of the current sheet is estimated to be larger than theoretical threshold of and thus ensures occurrence of tearing mode instability. The observations therefore suggest fragmented and turbulent magnetic reconnection occurring in the long stretching current sheet.
\end{abstract}


%\clearpage
\section{Introduction}
Magnetic reconnection, referring to dissipation and connectivity change of magnetic field, is capable of powering plasma heating, plasma motions, and particle acceleration in relativistic jets  \citep{bloom11}, accretion disks \citep{balbus98}, solar and stellar flares \citep{sturrock66}, and magnetospheres \citep{phan06}. In the past decades, abundant evidence for magnetic reconnection has been disclosed including in situ measurements near the Earth and remote sensing observations such as cusp-shaped flare loops \citep{masuda94}, inflows and downflows near the reconnection region \citep{yokoyama01,savage11,takasao12,liuw13,liurui13,xuezhike16}, double hard X-ray coronal sources \citep{sui03}, and changes of connectivity of coronal loops \citep{suyang13,yangshuhong15,lileping16_np}. 

Unfortunately, the specific processes involved in magnetic reconnection, in particular what occur in the reconnection region, remain mysterious. Theoretically, magnetic reconnection is believed to take place in a localised region, i.e., the so-called current sheet, that has enhanced resistivity \citep{priest14,yamada10}. In the Sweet-Parker model, the current sheet is limited to a long and thin region, in which the reconnection proceeds steadily but too slowly to interpret the real energy release rate. Through invoking slow-mode shocks extending from a shortened Sweet-Parker current sheet, the Petschek model is able to significantly boost the reconnection rate \citep{petschek64}. Nevertheless, the current sheet width in the Petschek model is of ion inertial scale, which can hardly match the detectable width in observations. Therefore, it was proposed that the current sheet can be fragmented into many magnetic islands by tearing mode instability \citep{furth63,shibata01} and develops turbulence to achieve the fast reconnection \citep{lazarian99}. However, such a picture has been short of direct observational evidence although documented by numerical simulations \citep{kowal09,shencc11,barta11} and indicated by various indirect observations such as simultaneous intermittent plasmoid ejections and hard X-ray/radio bursts \citep{asai04,nishizuka09,takasao16}, vortex above flare arcades \citep{mckenzie13,scott16}, and complex transition region line profiles with bright cores and broad wings \citep{innes15}. 

\begin{figure*} %%%%%%%%%%%%%%%%%% FIGURE 1
      \vspace{-0.0\textwidth}    % Shift back to the panel bottom
      \centerline{\hspace*{0.00\textwidth}
      \includegraphics[width=0.9\textwidth,clip=]{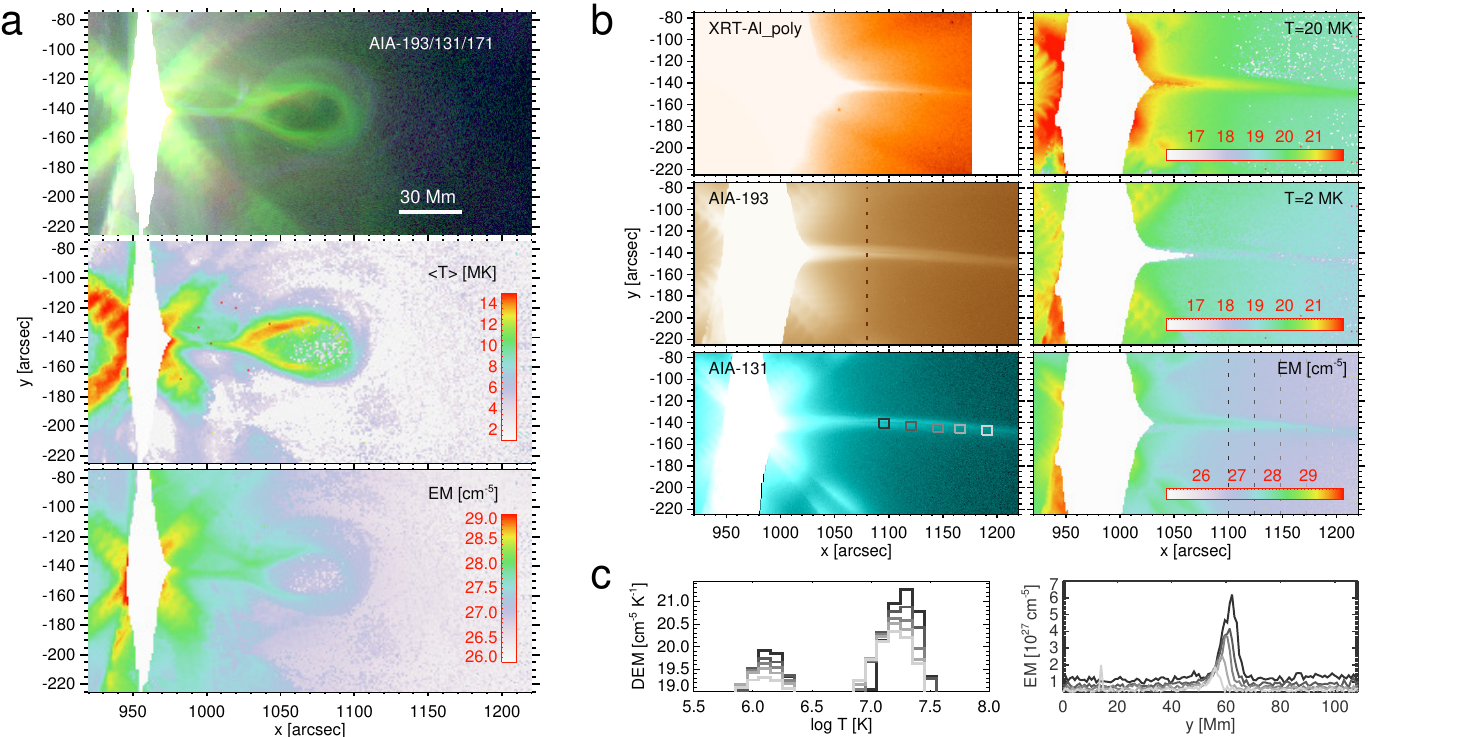}
      }
\caption{\textbf{Super-hot current sheet in the wake of an erupting bubble on 2017 September 10.} (a) Top: A composition of the AIA 193 {\AA} (red; temperature response peaks at $\sim$1.6 and 18 MK), 131 {\AA} (green; $\sim$0.4 and 11 MK), and 171 {\AA} (blue; $\sim$0.6 MK) images showing an erupting bubble and induced current sheet at 15:53 UT. Middle and bottom: DEM-weighted average temperature and total EM maps showing that the erupting bubble has a high temperature ($\sim$8 MK) but low density at its center. (b) Images at the Hinode-XRT Al-poly, SDO-AIA 193 {\AA} and 131 {\AA} passbands, DEM maps at the temperatures of 2 and 20 MK, and total EM map showing that the current sheet appears as a long and thin feature at 16:15 UT. The vertical dashed line in the AIA 193 {\AA} image indicates the location of the slit used to  construct the AIA time-distance plot in Figure \ref{f2}a. Note that, we do not calculate the average temperature and EM of the flare loops and cross-shaped structure as shown in panels a and b because the flux is saturated there (white region). (c) The DEM of the current sheet (left) at five specific regions (small boxes in panel b) and the total EM (right) along five dashed lines as shown in the EM map of panel b.} \label{f1}
\end{figure*}

In this study, we present a detailed analysis of a limb solar eruption on 2017 September 10 that produced an X8.2-class flare (SOL2017-09-10T16:06UT\footnote{http://sprg.ssl.berkeley.edu/~tohban/wiki/index.php}) and a fast coronal mass ejection (CME). In particular, the presence of a thin and long hot plasma sheet underneath an erupting CME fits perfectly into the current sheet structure, as predicted in the  theoretical model \citep{linjun00}, and the dynamic behaviours of the plasma within and around the current sheet provide direct and solid evidence of a turbulent and intermittent nature of magnetic reconnection.

\section{Instruments}
The data sets are primarily from Solar Dynamics Observatory \citep[SDO;][]{pesnell12}. The Atmospheric Imaging Assembly \citep[AIA;][]{lemen12} on board SDO images the solar corona with a spatial resolution of 0.6 arcsec per pixel and cadence of 12 seconds at 7 Extreme Ultraviolet (EUV) passbands. Here, we used the AIA data with a cadence of 24 seconds that have the normal exposure time. The data with a very short exposure time, in particular during the flare, usually have large uncertainties in intensity that can influence the differential emission measure (DEM) calculations and Fast Fourier Transform (FFT) analyses. The X-Ray Telescope \citep[XRT;][]{golub07} and EUV Imaging Spectrometer \citep[EIS;][]{culhane07} on board Hinode \citep{kosugi07} provide the X-ray images and EUV spectra in the wavelength ranges of 170--210 \AA~(short) and 250--290 \AA~(long) with a spectral resolution of 0.0223 \AA~pixel$^{-1}$, respectively. The Geostationary Operational Environmental Satellite (GOES) records the soft X-ray 1--8 {\AA} flux from the flare. In addition, the K-Cor instrument installed at the Mauna Loa Solar Observatory\footnote{https://www2.hao.ucar.edu/mlso/mlso-home-page} and the Large Angle and Spectrometric Coronagraph \citep[LASCO;][]{brueckner95} on board the Solar and Heliospheric Observatory (SOHO) observe the white-light images of the CME and its trailing current sheet. 

\section{Results}
\subsection{Hot Flux Rope and Induced Super-hot Current Sheet}
The early phase of the flare/CME eruption was fully captured by the AIA. At $\sim$15:35 UT, a filament is activated to rise up. After $\sim$15 min, it initiates the eruption of a nearby loop-like structure. Shortly afterwards, the loop-like structure quickly expands and escapes away from the solar surface. Simultaneously, the overlying field constraining the loop-like structure is stretched outwards. At $\sim$15:53 UT, the loop-like structure ascends to a height of 90 Mm and appears as a well defined bubble consisting of a ring-shaped envelop and a low emission cavity, both of which are visible at most EUV and X-ray passbands (top panel of Figure \ref{f1}a). An elongated bright structure connecting the bottom of the bubble and the top of the flare loops is observed. These features basically conform to the classic picture of eruptive flares \citep{sturrock66,shibata95,chen11_review}, in which the eruption of a twisted magnetic flux rope leaves behind a long and narrow current sheet \citep{linjun00}. The bubble is most likely an edge-on manifestation of the forming flux rope as its axis is mostly along the line-of-sight \citep{cheng11_fluxrope}. The differential emission measure (DEM) analyses show that the cavity of the bubble has a low emission measure (EM$\sim$$10^{26}$ cm$^{-5}$), though the temperature is relatively high ($\sim$10 MK). By contrast, the bubble envelope (or the ring) and the current sheet have a much higher emission measure ($\sim$$10^{27.5}$ cm$^{-5}$) and an even higher temperature ($\sim$13 MK). Such a temperature structure highly resembles the numerical results of the erupting flux rope energised by the reconnection in its trailing current sheet \citep{mei12}.

\begin{figure*} %%%%%FIGURE 2%%%%%%%%%%%
\vspace{0.0\textwidth}    % Shift back to the panel bottom
     \centerline{\includegraphics[width=0.4\textwidth]{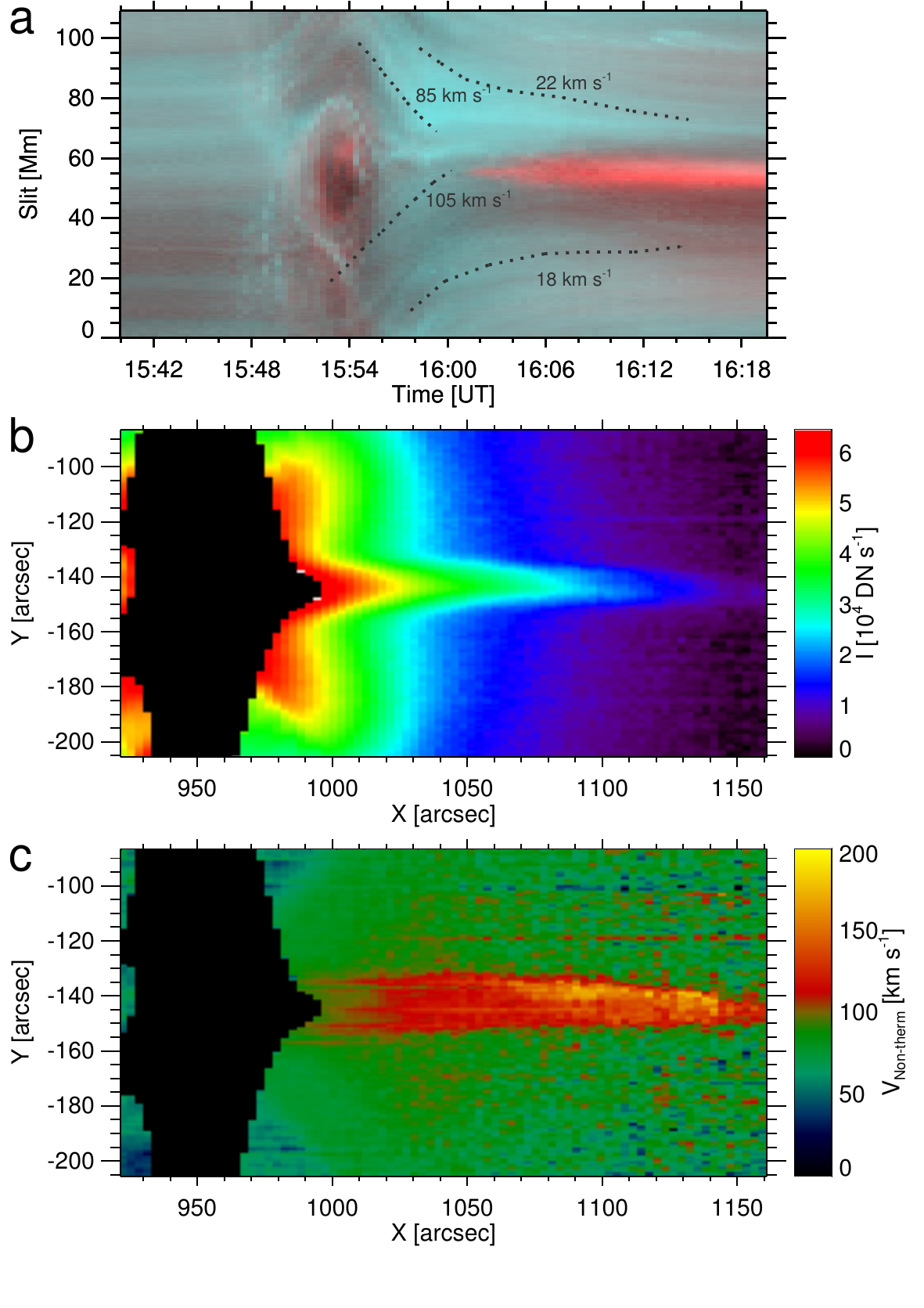}
     \hspace{-0.6\textwidth} \vspace{-0.0\textwidth}
     \centerline{\includegraphics[width=0.4\textwidth,clip=]{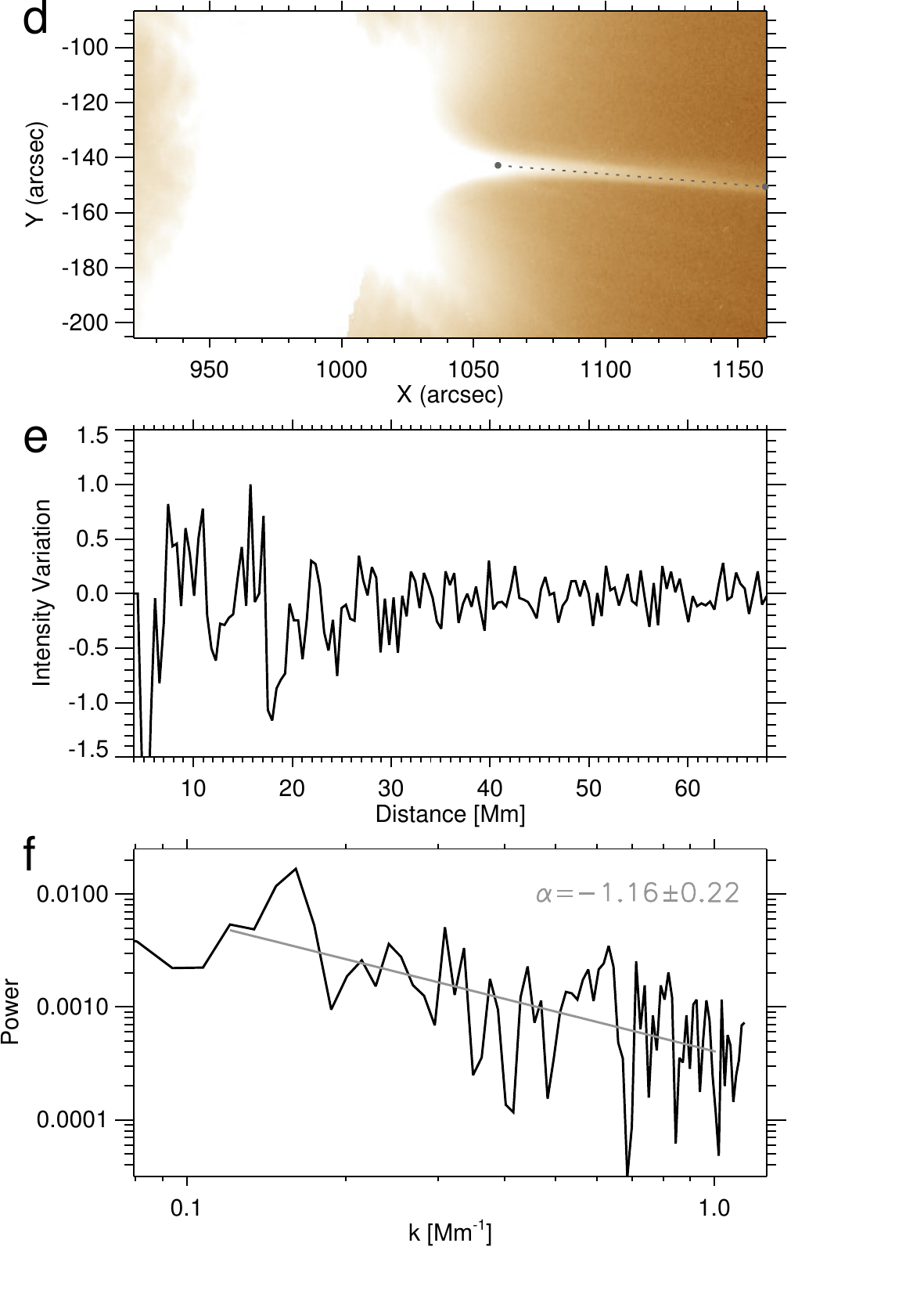}\hspace{-0.6\textwidth} \vspace{-0.0\textwidth}}
     \hspace{0.0\textwidth}}
     \vspace{0.0\textwidth}
%{\includegraphics[width=1\textwidth,clip=]{f2_1.pdf}\hspace{-0.0\textwidth} \vspace{-0.0\textwidth}}
               \vspace{0.0\textwidth}
\caption{\textbf{Evidence of inflows and turbulence}. (a) The time-distance plot of the AIA 193 {\AA} (red) and 171 {\AA} (cyan) composited images showing the converging inflows, whose trajectories are tracked by the dashed lines. Their velocities range from 20 to 100 km s$^{-1}$. (b and c) Intensity and non-thermal velocity field of the EIS Fe XXIV 192.03 {\AA} line. The imaging spectra are obtained by the EIS slit scanning the current sheet region from 16:09 UT to 16:18 UT. (d) The AIA 193 {\AA} image shows the current sheet structure at 17:10 UT as indicated by the dotted line. (e) The spatial distribution of the normalised AIA 193 {\AA} intensity along the dotted line in panel d. The intensity is detrended with a moving average of 10 Mm to indicate the fast-varying structures. (f) The power spectrum of the detrended intensity variation as shown in panel (e) in spatial frequency domain. The fitting spectral index $\alpha$ in the range of 1--10 Mm is --1.16$\pm$0.22.} \label{f2}
\end{figure*}

As the bubble escapes from the lower corona, the current sheet is further heated and extended. Its lower end ascends to a height of at least $\sim$100 Mm around the flare peak time of $\sim$16:15~UT (Figure \ref{f1}b). The EM maps at different temperatures document that the extended current sheet mainly contains high temperature plasma (Figure \ref{f1}b), which is also confirmed by the EIS Fe XXIV 192.03 and 255.11 {\AA} lines (with the formation temperature of $\sim$18 MK). The plasma therein is primarily distributed near the temperature of 20 MK with the total EM of 1--5$\times$10$^{27}$ cm$^{-5}$ (Figure \ref{f1}c), which is similar to the temperature of supra-arcade downflows that are frequently observed when the current sheet is observed face-on \citep{Hanneman14}. The corresponding density is calculated to be $\sim$0.6--1.3$\times$10$^{9}$ cm$^{-3}$ assuming a depth of 30 Mm (the size of the bubble) at the height of $\sim$100--200 Mm. Based on the distribution of the total EM along the direction perpendicular to the current sheet, the average width of the current sheet is estimated to be $\sim$10 Mm at the height of $\sim$150 Mm (Figure \ref{f1}c), slightly larger than the width estimated by \citet{savage10}.

\subsection{Fragmented and Turbulent Current Sheet} 
\begin{figure*} %%%%%%%%%%%%%%%%% FIGURE 3
\vspace{0.0\textwidth}
\centerline{\includegraphics[width=0.9\textwidth,clip=]{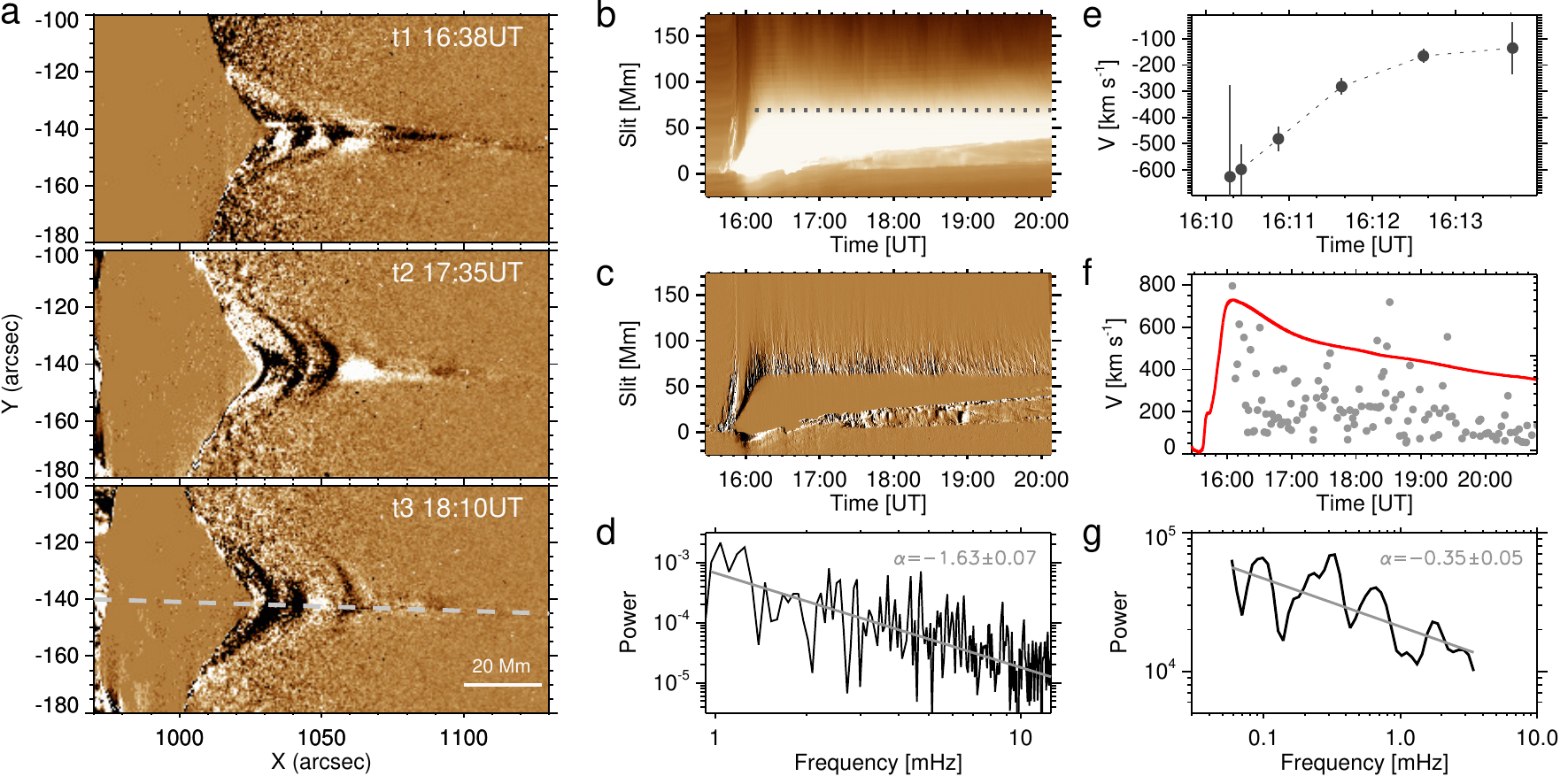}\hspace{0.0\textwidth} \vspace{0.0\textwidth}}
\caption{\textbf{Intermittency and velocity diversity of the sunward outflow jets.} (a) The AIA 193 {\AA} running difference images (the time difference is 24 seconds) at 16:38 UT (top), 17:35 UT (middle), and 18:10 UT (bottom) displaying the reconnection outflow jets (white features) expelled from the lower end of the current sheet. (b) and (c) Time-distance plots of the original images and running difference images at the AIA 193 {\AA} passband along the direction shown by the dashed line in panel a. The white spicule-like features after $\sim$16:00 UT as shown in panel c clearly display the sunward downflows. (d) The power spectrum of the normalised AIA 193 {\AA} intensity in the outflow region as a function of frequency. The location is indicated by the dashed line in panel b. Similar to Figure \ref{f2}f, the intensity is also detrended with a moving average of 60 min in order to show the high frequency component. The spectral index ($\alpha$) derived by linear fitting to the data in the range of 1--15 mHz is --1.63$\pm$0.07. (e) The velocity evolution for one reconnection outflow jet. The errors of the velocities (vertical bars) are from the measurement uncertainties in height ($\sim$1.7 Mm). (f) Scatter plot of the initial velocities of the outflow jets as a function of time. The GOES soft X-ray 1--8 {\AA} flux is also plotted for comparison. (g) The power spectrum of the initial velocities as a function of frequency. The spectral index $\alpha$ is --0.35$\pm$0.05.}\label{f3}
(Animations of Figure \ref{f2}a is available.)
\end{figure*}

The EUV 171 {\AA} observations disclose that the cool plasma ($\sim$1 MK) on both sides converges into the current sheet (Figure \ref{f2}a). Shortly afterwards, the plasma is strikingly heated and becomes visible in the AIA higher temperature passbands such as 193 {\AA} and 131 {\AA} (10--20 MK). The average velocity of the converging motion is $\sim$100 km s$^{-1}$ in the early phase (15:55--16:00 UT) and subsequently decreases to $\sim$20 km s$^{-1}$, similar to previous estimations \citep{liuw13,zhucm16,liying17,wangjt17}. The initial and faster inflows are possibly driven by the restoring force of the magnetic field, which was pushed aside by the erupting bubble before $\sim$15:54 UT. Besides the plasma heating, the Fe XXIV 192.03 {\AA} line also displays a significant non-thermal broadening in the current sheet. The line width implies a non-thermal velocity of $\sim$100--150 km s$^{-1}$ after subtracting the thermal velocity corresponding to a formation temperature of 18 MK (Figure \ref{f2}b and \ref{f2}c). Such large non-thermal velocity strongly indicates the existence of turbulent motions in the current sheet \citep[also see][]{ciaravella08,Doschek14,warren17}. It is also supported by the fact that the 193 {\AA} intensity variation along the current sheet presents a fluctuation, which shows a power-law distribution in spatial frequency domain after Fast Fourier Transform (FFT) (Figure \ref{f3}d and \ref{f3}e). The spectral index is estimated to be --1.16$\pm$0.22 (Figure \ref{f3}f).

The turbulent current sheet indicates that the sunward reconnection outflow jets, probably corresponding to magnetic islands expelled from the lower end of the current sheet, will show a power-law behaviour. Figure \ref{f3}a and attached movie clearly show that the jets are intermittently shot out during the reconnection process. Each jet has an ``Eiffel Tower" shape initially. Within a period of 2--5 min, probably driven by magnetic tension \citep{forbes96,priest02}, each jet gradually becomes to be cusp-shaped, and then continuously shrinks to a flare loop. The 193 {\AA} intensities in the outflow regions also present intermittent fluctuations (Figure \ref{f3}b and \ref{f3}c). The FFT analysis shows that the temporal variation of the intensity (e.g., along the dashed line in Figure \ref{f3}b) does obey the power law distribution with a spectral index around --1.60 (Figure \ref{f3}d), very close to the spectral index of the turbulent current sheet \citep[e.g.,][]{barta11,shencc11}. It confirms our conjecture that the current sheet has been fragmented into different scaled structures, strongly suggestive of the existence of turbulence, with which the outflow jets are widely distributed in energies and sizes. Further evidence for a fragmented and turbulent reconnection is that the intensity variations at the other flaring regions also present the power law distribution with spectral indices ranging from --1.2 to --1.8, quite different from that in the quiescent and pre-flare regions (see Figure \ref{f3_193_131}--\ref{f3_131_1} in Appendix). It is worthy of noticing that supra-arcade downflows may directly correspond to the sunward outflow jets \citep{mckenzie02,savage11,reeves15} or be structures caused by Rayleigh Taylor instabilities in the outflow region \citep{guolj14}.

We find that the velocity of the sunward outflow jets also presents a dispersed distribution. The heights of the jets are measured through manually tracking their trajectories (as shown by Figure \ref{fs_track} in Appendix). Almost all outflow jets have a large initial velocity but quickly slow down (Figure \ref{f3}e). The initial velocities are diversely distributed, ranging from 100 to 800 km s$^{-1}$ (Figure \ref{f3}f and Figure \ref{fs_track}), even in a short time period (16:00--16:30~UT), similar to previous estimations \citep{savage11}. Interestingly, the FFT analysis indicates that the initial velocities also have a power law spectrum with a spectral index of --0.35 (Figure \ref{f3}g). It implies that the reconnection that drives the outflow jets proceeds with a varying reconnection rate, probably modulated by turbulence. Taking an average value (20 km s$^{-1}$) of the inflow velocities near the flare peak time (16:00--16:15 UT), the reconnection rate (the ratio of the inflow velocity to the outflow velocity) is estimated to range from 0.003 to 0.2. If the current sheet is fragmented into magnetic islands of different sizes \citep{shibata01}, different reconnection rates and thus different kinetic energies of the outflow jets can be achieved. Moreover, the decelerations also have a wide distribution with its maximum up to 2000 m s$^{-2}$ (Figure \ref{fs_track}), indicating that the upward magnetic pressure gradient force also varies with time that resists the downward magnetic tension and the Sun's gravity.

\begin{figure*} %%%%%FIGURE 2%%%%%%%%%%%
\vspace{0.0\textwidth}
\centerline{\includegraphics[width=0.44\textwidth,clip=]{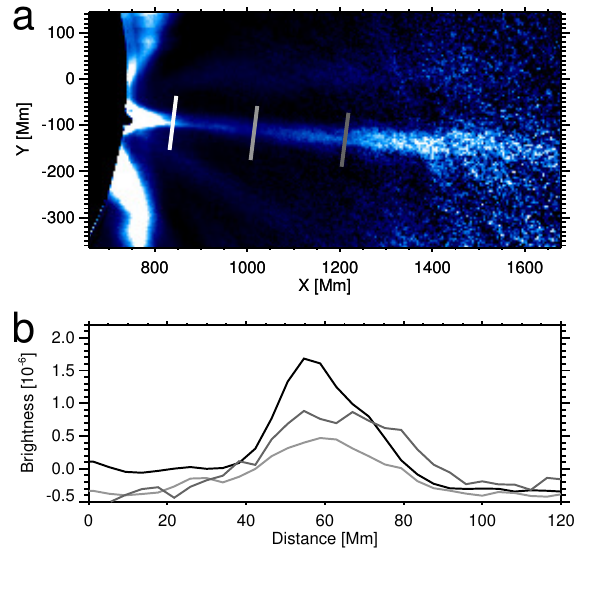}\hspace{0.46\textwidth} \vspace{-0.445\textwidth}}
\centerline{\includegraphics[width=0.465\textwidth,clip=]{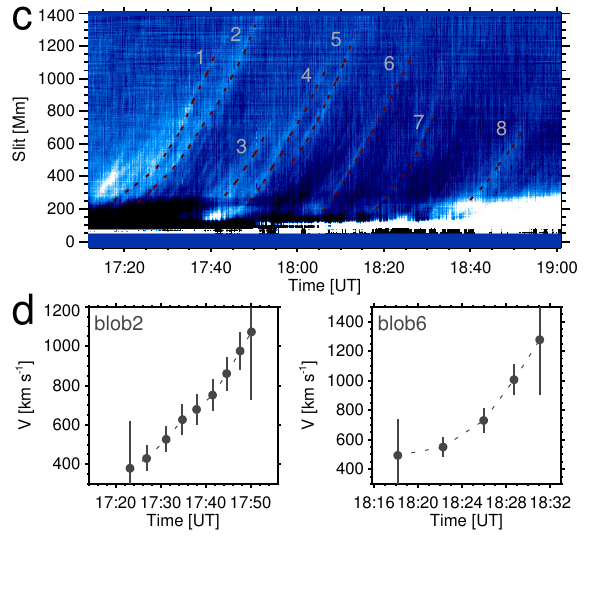}\hspace{-0.46\textwidth} \vspace{-0.0\textwidth}}
\caption{\textbf{Extended current sheet and intermittent anti-sunward moving blobs.} (a) White-light K image normalised with a radially graded filter observed by the Mauna Loa Solar Observatory showing the extended current sheet at 17:15 UT. (b) The brightness distributions of the current sheet along three perpendicular slits (in panel a). (c) Time-distance plots of the running difference images at white-light K band along the direction of the current sheet. The dashed lines denote the trajectories of eight anti-sunward moving blobs. (d) The velocity evolution for blob 2 and blob 6. The errors of the velocities (vertical bars) are from the measurement uncertainties in height ($\sim$23 Mm)} \label{f4}
\end{figure*}

\subsection{Largely Extended White-light Current Sheet}
The K-Cor instrument of the Mauna Loa Solar Observatory provides the white-light images of the largely extended current sheet at its later phase. At 17:12 UT, the lower end of the current sheet is seen to joint the tip of the cusp-shaped flare loops and is located at a height of $\sim$140 Mm (Figure \ref{f4}a), similar to the value measured in the EUV data. The apparent width of the current sheet is $\sim$25 Mm, and the lower limit of the apparent length is 400 Mm (Figure \ref{f4}b). It corresponds to a maximal reconnection rate of $\sim$0.06, which, similar to the previous estimations \citep{savage10,ling14,seaton17}, is still smaller than the maximum value derived above. In fact, the original current sheet could be fragmented into magnetic islands due to tearing mode instability. Then, the real length of magnetic islands involved in each elementary reconnection process could be much smaller. This is proved by the fact that the length-to-width ratio ($>$16) of the current sheet is much larger than the theoretical threshold of tearing mode instability (2$\pi$) \citep{furth63}. The high-speed anti-sunward moving blobs also provide strong evidence for existence of magnetic islands. Figure \ref{f4}c shows that the blobs are intermittently formed in the current sheet at the height of $\sim$200 Mm. The initial velocities are $\sim$400 km s$^{-1}$ and then gradually increase. Note that, the width of the current sheet derived in the K corona is about 2.5 times larger than that in the EUV passbands. However, both are still much smaller than the values measured previously in the LASCO/C2 white-light coronagraph ($\sim$100 Mm at 2 $R_\odot$ \citep{linjun05,linj09,linj15,ciaravella08}). Interestingly, the EUV current sheet is found to be located in the middle of the white-light sheet, implying that the former is closer to the dissipation layer and thus has a higher temperature. Theoretically, the width of the diffusion layer is only tens of km for Petschek-type magnetic reconnection with an anomalous resistivity. However, in observations, the apparent width of the current sheet can be seriously widened by turbulence, as well as slow-mode shock compression and projection effects \citep{ciaravella08,linj15}.

\section{Summary and Discussions}
In the models of flux-rope-induced CME/flare eruptions \citep{shibata95,chen11_review}, a pre-existing flux rope escapes away from the solar surface due to loss of equilibrium \citep{linjun00}, leading to the formation of a CME and a flare at almost the same time \citep{zhang01,cheng11_fluxrope}. Magnetic reconnection acts as strong coupling between the CME and the flare as indicated by the simultaneity between the evolution of the CME velocity and the variation of the flare emission \citep{zhang01}. The linear bright feature in the wake of the erupting flux rope has been argued to be the induced current sheet, where electric current is enhanced and magnetic field is dissipated \citep{linj15}. Previous observations of the current sheet are mostly limited by the wavelength window that only responds to relatively narrow and low temperatures and/or the field of view that is not large enough \citep{linjun05,linj09,ciaravella03,ciaravella08,savage10,ling14,seaton17}. Therefore, studies based on these observations are mostly speculative in particular on the origin of the current sheet and its relation to the CME and flare. Moreover, the observations in those works could not provide further information on the detailed physical processes occurring in the current sheet, therefore it has seldom been addressed what kind of reconnection it is. 

In this study, we present a solar limb eruption event, which displays a distinct picture of the CME/flare eruption with unprecedented clarity. Observations with a continuous field of view from 1 to 30 $R_\odot$ and multi-wavelengths including the white-light, EUV, and X-rays enable us to reveal the origin of and specific processes involved in magnetic reconnection. We successfully detect almost all ingredients predicted by models during a single eruption including the erupting hot flux rope, super-hot current sheet, cusp-shaped flare loops, inflows, and high-speed sunward and anti-sunward outflow jets, some of which have been detected in previous observations \citep{savage10,ling14,seaton17,yanxl18_cs,liuwei18}. The high temperature of the flux rope envelope and the cusp-shaped flare loops probably originates from the collision of the outflow jets with the local dense plasma and/or the direct heating by slow-mode shocks at both ends of the current sheet \citep{liuw13}. The high temperature of the current sheet, however, requires a local heating by magnetic energy dissipation inside the current sheet itself. 

The turbulent behaviour of energy release in the current sheet is also revealed. A high Lundquist number, suggested by a large length-to-width ratio ($>$16) of the current sheet, leads to the generation of magnetic islands due to tearing mode instability \citep{furth63}, which subsequently appear as intermittent sunward outflow jets and anti-sunward moving blobs when shot out of the current sheet. Simultaneously, the turbulence develops in the current sheet \citep{strauss88,lazarian99}. On the one hand, its effect helps achieve anomalous resistivity to boost magnetic dissipation rate. On the other hand, it may mediate the formation of magnetic islands with their size and energy presenting a power law distribution. This process finally makes the intensity and velocity of the sunward outflow jets exhibit a power law distribution. In particular, the spectral index of the former is found to vary from --1.2 to --1.8, which suggests that the turbulence mediate the reconnection process in the current sheet, resulting the formation of different scaled magnetic islands, consistent with previous numerical results \citep{kowal09,shencc11,barta11}. The deviation from the fully developed isotropic turbulence (with a Kolmogorov turbulence spectral index of --5/3) may be due to the role of magnetic field. The significant non-thermal motions shown in the Fe XXIV line also evidence the existence of turbulence. In summary, these observations show that the magnetic reconnection, at least in solar eruptions, does not proceed uniformly in space and time. Instead, the current sheet should be composed of fragmented structures, in which magnetic reconnection dissipates magnetic energy in a turbulent way \citep{kontar17} to heat the plasma and drive the outflow jets. 

\acknowledgements We are cordially grateful to five anonymous referees for their very meaningful comments and suggestions.  We also thank Jun Lin, Zongjun Ning, Dong Li, Yu Dai and Jinsong Zhao for their helpful discussions. AIA data are courtesy of NASA/SDO, a mission of NASA's Living With a Star Program. XRT and EIS data are courtesy of Hinode, a Japanese mission developed and launched by ISAS/JAXA. KCOR data are courtesy of Mauna Loa Solar Observatory operated by the High Altitude Observatory. X.C., Y.L., L.F.W., M.D.D., \& P.F.C. are supported by NSFC through grants 11722325, 11733003, 11790303, 11790300 and by Jiangsu NSF through grants BK20170011. X.C. is also supported by ``Dengfeng B" program of Nanjing University. Y.L. is also supported by CAS Pioneer Hundred Talents Program. J.Z. is supported by US NSF.

\begin{center}
\textbf{APPENDIX}
\end{center}

%\begin{center}
%\textbf{Methodology and Uncertainties}
%\end{center}

\begin{figure} %%%%%%%%%%%%%%%%
\vspace{0.0\textwidth}
\centerline{\includegraphics[width=0.5\textwidth,clip=]{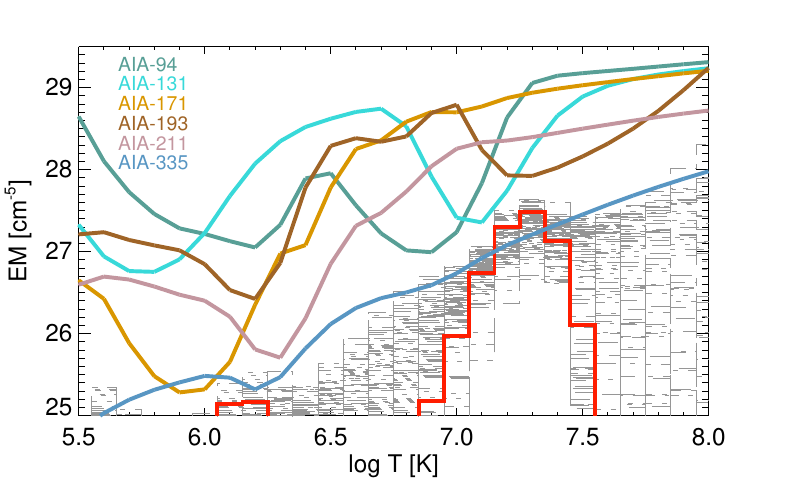}\hspace{0.0\textwidth} \vspace{-0.0\textwidth}}
\caption{Loci curves of the emission in a small region in the current sheet, whose position is shown by the black box in Figure \ref{f1}c. The red line is the best-fitting of the EM distribution derived by ``xrt\b{ }dem\b{ }iterative2.pro". The gray dashed lines represent 100 MC solutions. The EM is calculated by Equation (2) in each temperature bin.} \label{fs_dem}
\end{figure}

\begin{figure*} %%%%%%%%%%%%%%%%
\vspace{0.0\textwidth}
\centerline{\includegraphics[width=0.9\textwidth,clip=]{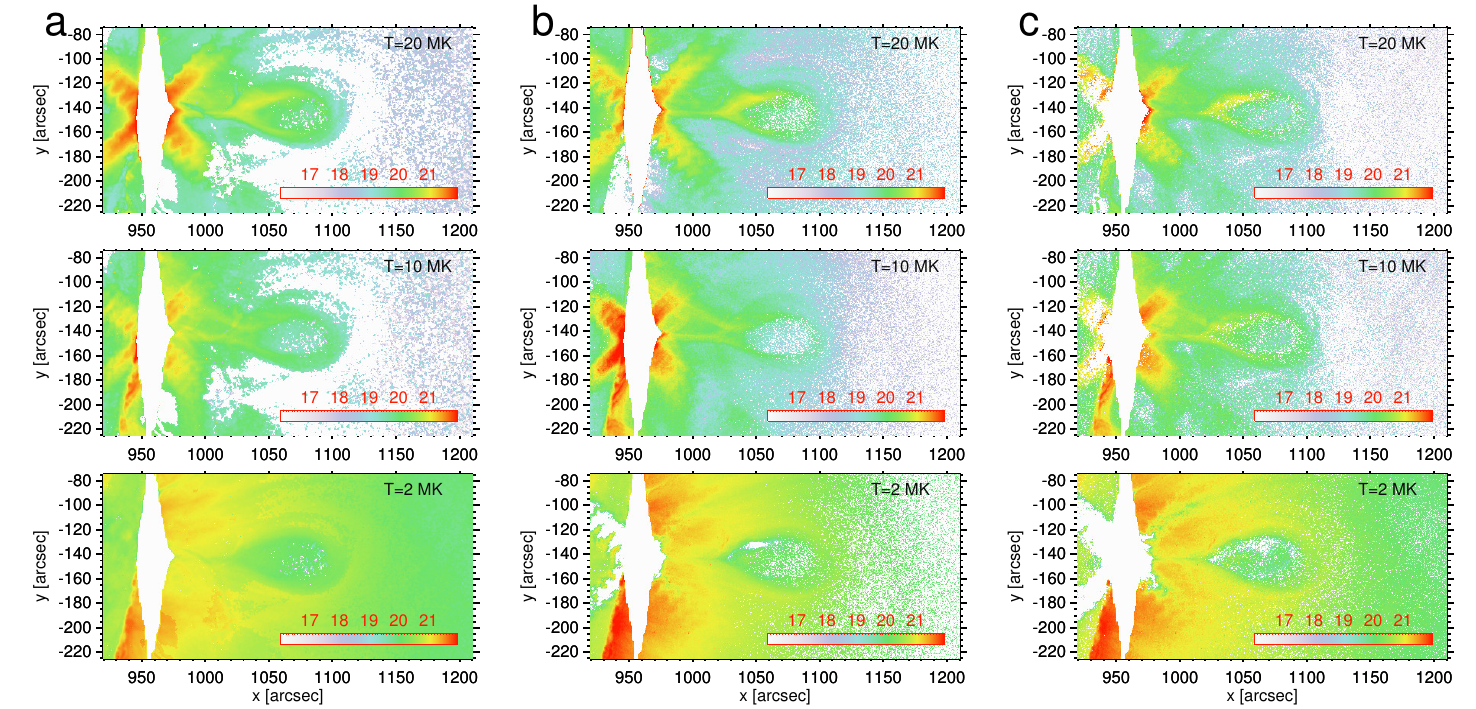}\hspace{0.0\textwidth} \vspace{-0.0\textwidth}}
\caption{DEM maps of the erupting hot bubble at the temperatures of 20 MK, 10 MK and 2 MK derived by the methods of Weber, M.~A. (a), Hannah, I. G. (b), and Cheung, M.~C.~M. (c), respectively.} \label{fs_dem1}
\end{figure*}

\begin{figure*} %%%%%%%%%%%%%%%%
\vspace{0.0\textwidth}
\centerline{\includegraphics[width=0.9\textwidth,clip=]{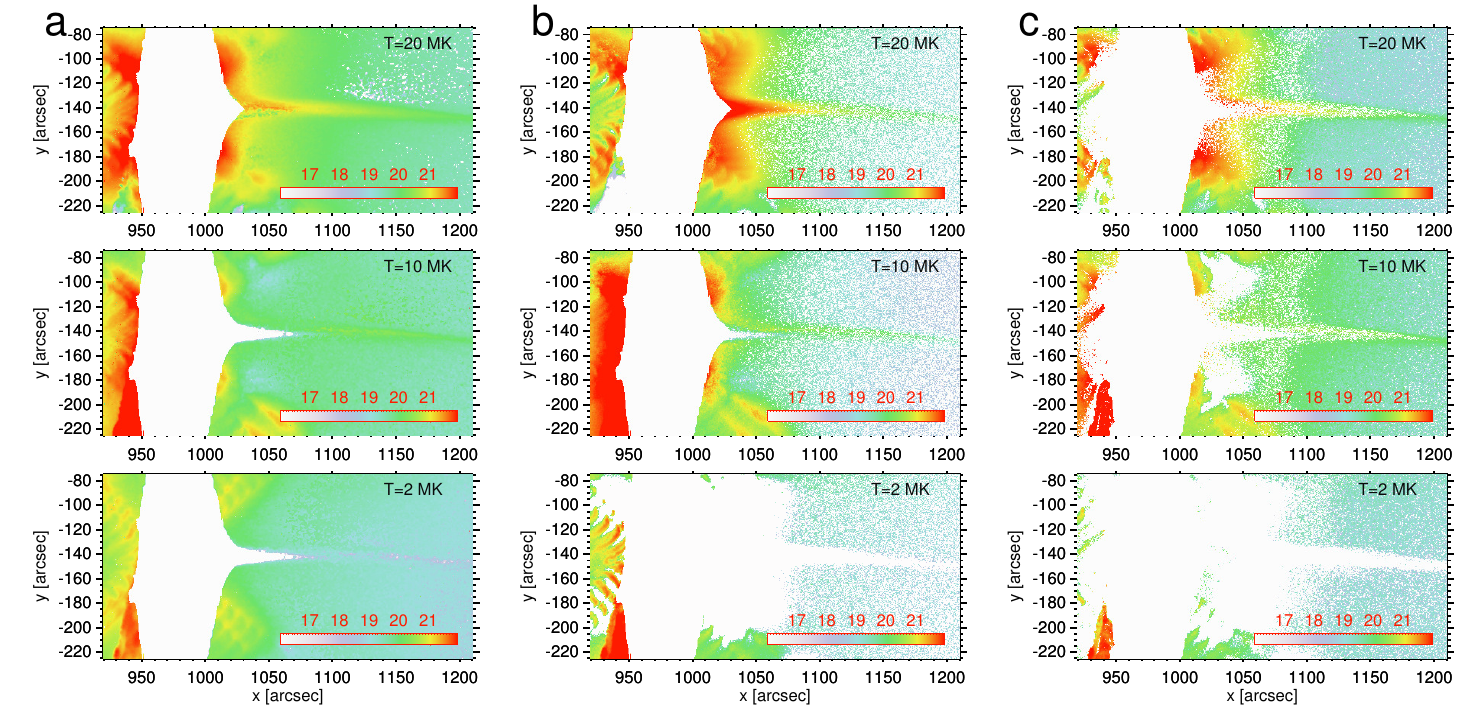}\hspace{0.0\textwidth} \vspace{-0.0\textwidth}}
\caption{Same as Figure \ref{fs_dem1} but for the current sheet.} \label{fs_dem2}
\end{figure*}

\begin{figure*} %%%%%%%%%%%%%%%%
\vspace{0.0\textwidth}
\centerline{\includegraphics[width=0.9\textwidth,clip=]{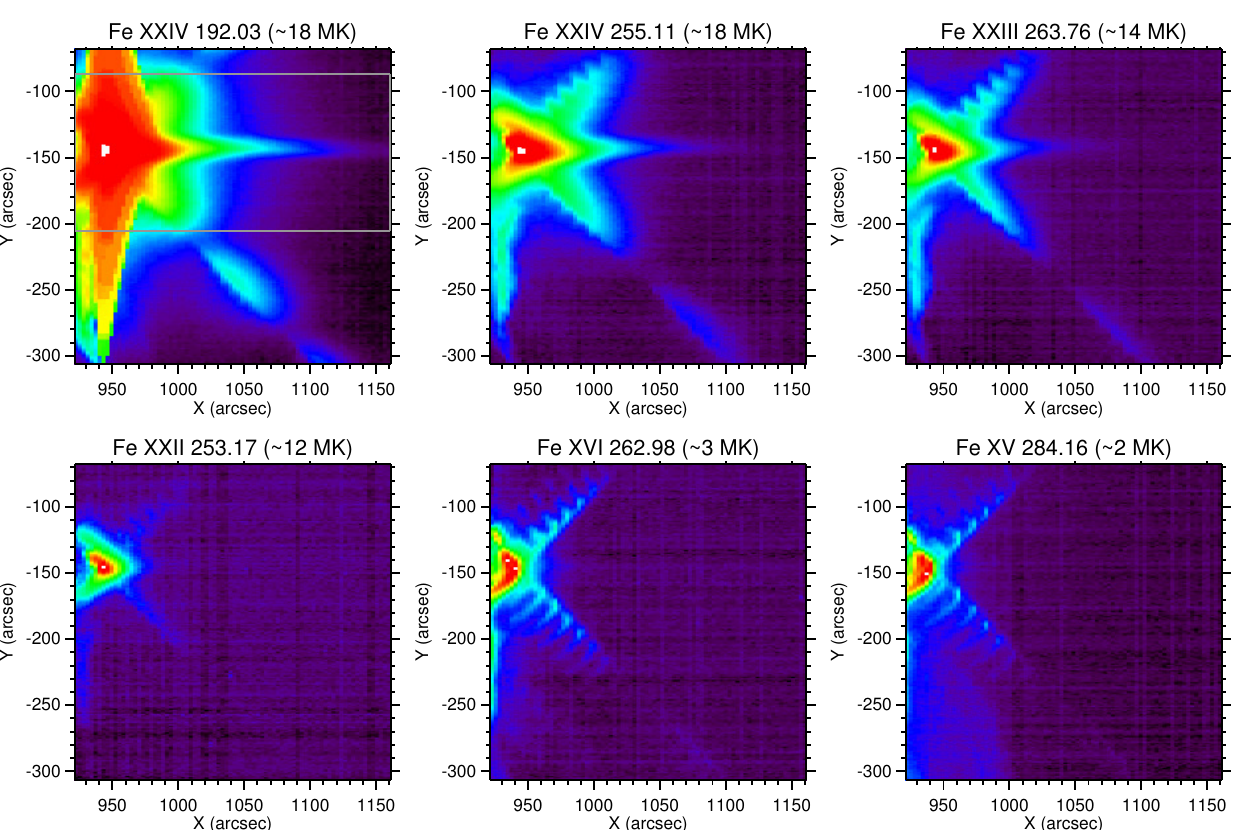}\hspace{0.0\textwidth} \vspace{-0.0\textwidth}}
\caption{The Fe XXIV192.03 {\AA} ($\sim$18 MK), 255.11 {\AA} ($\sim$18 MK), Fe XXIII 263.76 {\AA} ($\sim$14 MK), Fe XXII 253.17 {\AA} ($\sim$12 MK), Fe XVI 262.98 {\AA} ($\sim$3 MK) and Fe XV 284.16 {\AA} ($\sim$2 MK) line spectra showing the visibility and invisibility of the current sheet. The box in the upper-left panel indicates the region where intensities and non-thermal velocities of the Fe XXIV192.03 {\AA} line are shown in Figure 2.} \label{fs_eis}
\end{figure*}

\begin{figure*} %%%%%%%%%%%%%%%%
\vspace{0.0\textwidth}
\centerline{\includegraphics[width=0.9\textwidth,clip=]{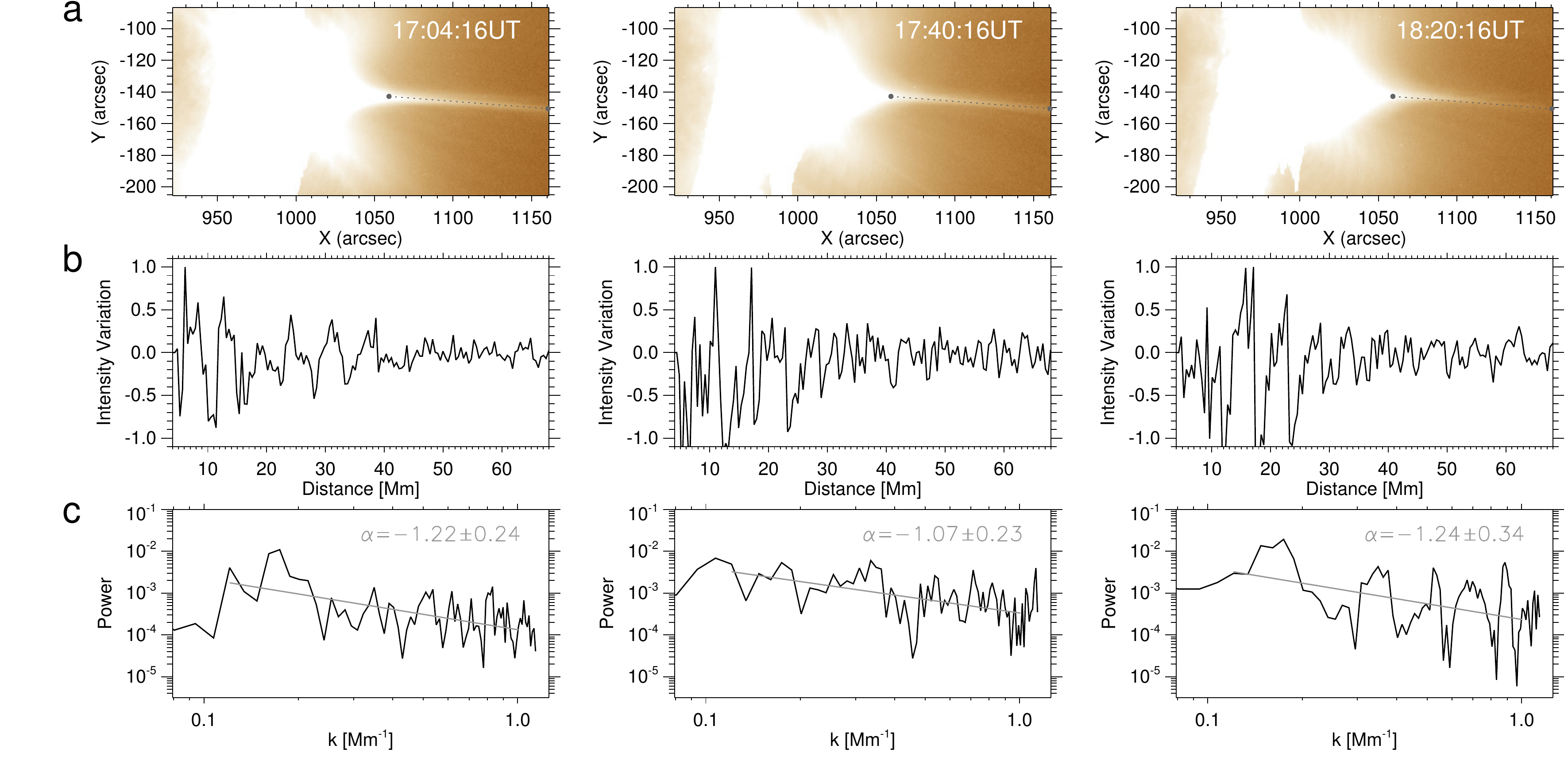}\hspace{0.0\textwidth} \vspace{-0.0\textwidth}}
\caption{(a) The AIA 193 {\AA} images showing the evolution of the current sheet. (b) The spatial variation of the normalised AIA 193 {\AA} intensity along the current sheet indicated by the dotted line in panel a. (c) The power spectrum of the intensity variation as a function of spatial frequency with the oblique lines showing the power-law fitting to the range of 1--8 Mm.} \label{f_spectral_193}
\end{figure*}
 
\begin{figure*} %%%%%%%%%%%%%%%%
\vspace{0.0\textwidth}
\centerline{\includegraphics[width=0.9\textwidth,clip=]{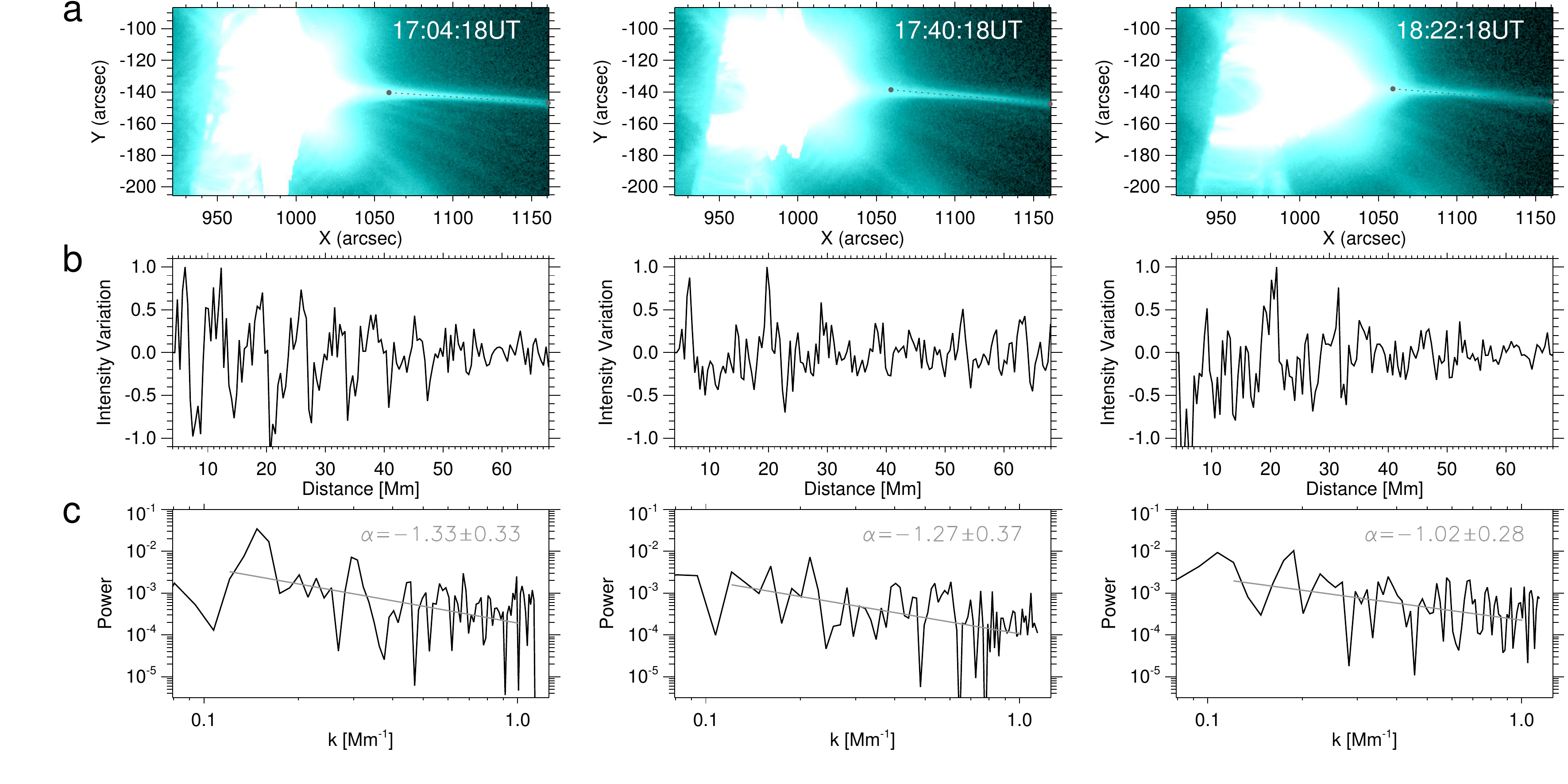}\hspace{0.0\textwidth} \vspace{-0.0\textwidth}}
\caption{Same as Figure \ref{f_spectral_193} but for the AIA 131 {\AA} passband.} \label{f_spectral_131}
\end{figure*}

\textbf{DEM reconstruction:} 
The DEM is resolved through six AIA passbands including 94 {\AA} (Fe X, $\sim$1.1 MK; Fe XVIII, $\sim$7.1 MK), 131 {\AA} (Fe VIII, $\sim$0.4 MK; Fe XXI, $\sim$11 MK), 171 {\AA} (Fe IX, $\sim$0.6 MK), 193 {\AA} (Fe XII, $\sim$1.6 MK; Fe XXIV, $\sim$18 MK), 211 {\AA} (Fe XIV, $\sim$2.0 MK), and 335 {\AA} (Fe XVI, $\sim$2.5 MK). The code ``xrt\b{ }dem\b{ }iterative2.pro" in Solar SoftWare \citep[SSW;][]{Freeland98}, originally proposed by \cite{weber04} and later modified by \cite{cheng12_dem}, is used for reconstructing the DEM. The inputs are observed intensity $I_{i}$ and the temperature response function $R_{i}(T)$ of the passband $i$. $I_{i}$ is written as:
 \begin{equation}
 {I_{i}}  =  \int DEM\times R_i (T) \mathrm{d}T + \delta I_{i},
 \end{equation}
where $DEM$ denotes the plasma DEM, and $\delta I_{i}$ is the uncertainty in the intensity $I_{i}$. The temperature range for doing the inversion is set as 5.5$\leq$ log${T}\leq$ 8.0. The EM is calculated as:
 \begin{equation}
 EM= \int DEM dT,\\
 \end{equation}
where the temperature range of integration is set to be 7.0$\leq$ log${T}\leq$ 7.5. We performed 100 Monte Carlo (MC) solutions through adding a random noise (within the errors of observed intensities) to the intensity $I_{i}$ and then rerunning the routine. The results show that 100 MC solutions are converged in the range of 7.0$\leq$ log${T}\leq$ 7.5 (Figure \ref{fs_dem}). The density $n$ in the current sheet is obtained by:
\begin{equation}
n= \sqrt{EM/l},\\
\end{equation}
where $l$ is the depth of the current sheet.

We also take advantage of other two inversion methods independently developed by \cite{hannah12} and \cite{cheung15_dem}, respectively. It is found that, the three methods give very similar results. The erupting bubble, in particular its envelope, primarily contains high temperature plasma (Figure \ref{fs_dem1}), while the background and foreground contribute some cool plasma emission. As for the current sheet, the results from the different methods are also consistent with each other, which all present a super-hot ingredient and absence of cool plasma in the current sheet (Figure \ref{fs_dem2}). It is noticed that some discrepancies also exist. The code ``xrt\b{ }dem\b{ }iterative2.pro" is able to reconstruct the super-hot current sheet with a pretty good clarity. However, in the region outside of the current sheet, it may overestimate the DEM values compared with the other two codes. Nevertheless, we do not think that it influences our results, at least qualitatively. The results are also consistent with \cite{warren17}, who did the DEM inversion via the combination of AIA and EIS data and also found that the plasma in the current sheet has temperatures of about 20 MK and distributes in a relatively narrow temperature range. 
  
The uncertainties in the DEM results come mainly from the uncertainties in the observed intensities, which are obtained by the routine ``aia\b{ }bp\b{ }estimate\b{ }error.pro" in SSW. The uncertainties of the intensities are a result of the uncertainties in the temperature response functions of AIA including non-ionization equilibrium effects \citep{imada11}, non-thermal populations of electrons, modifications of dielectronic recombination rates \citep{summers74,badnell03}, radiative transfer effects \citep{judge10}, and even unknown filling factor. After considering these effects, an uncertainty lower limit of $\sim$20\% for $R_{i}(T)$ is derived \citep{judge10} and thus can not significantly influence the results \citep{cheng12_dem}.

\begin{figure*} %%%%%%%%%%%%%%%%
\vspace{-0.1\textwidth}
\centerline{\includegraphics[width=0.85\textwidth,clip=]{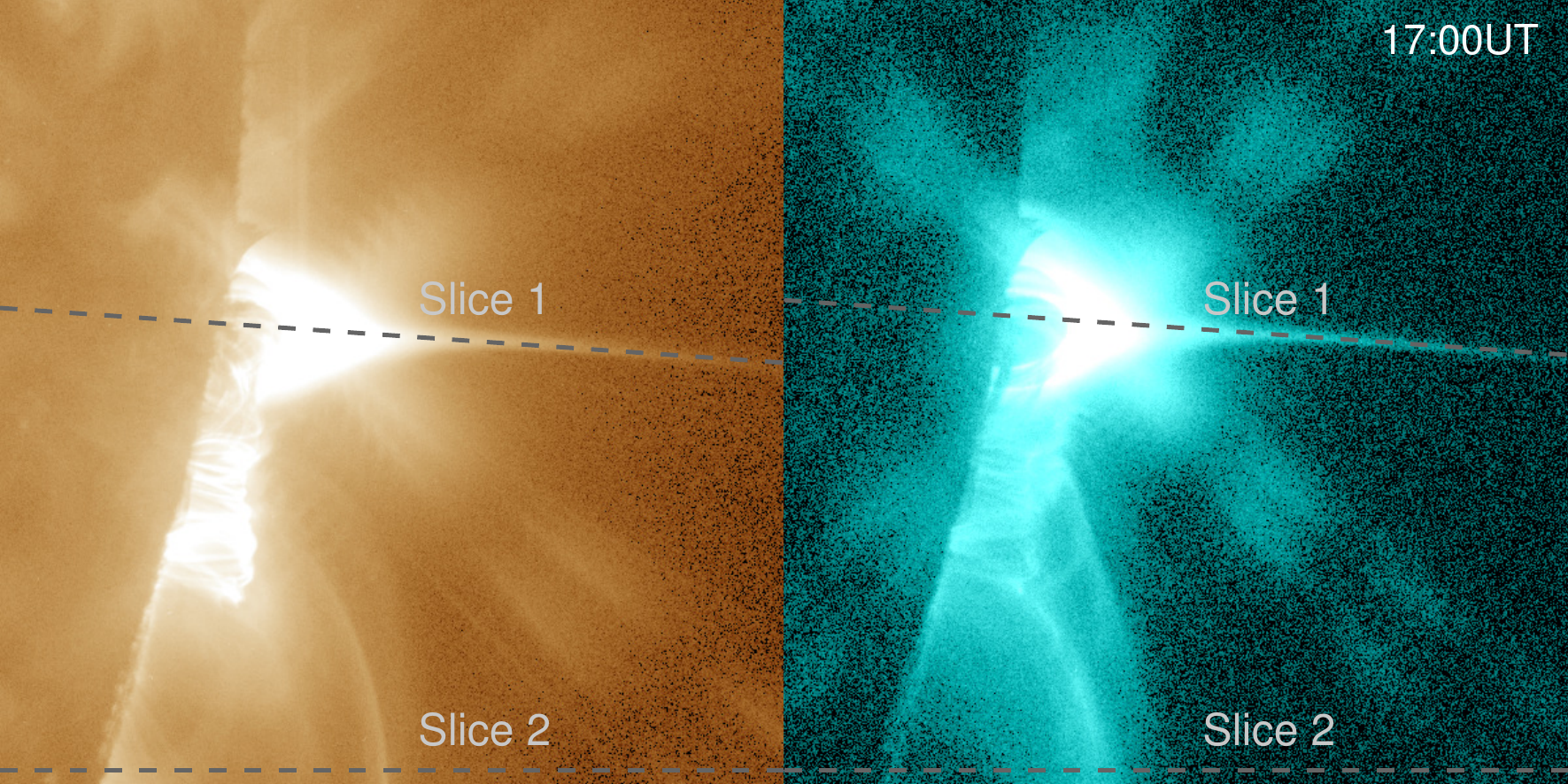}\hspace{0.0\textwidth} \vspace{-0.0\textwidth}}
\caption{The AIA 193 {\AA} and 131 {\AA} images at 17:00 UT showing the post-flare loops and the current sheet. The slice 1 and slice 2 are used for creating the time-distance plots in Figure \ref{f3_193}a-\ref{f3_131_1}a.} \label{f3_193_131}
\end{figure*}

\textbf{Spectroscopic Analyses:} 
The EIS data are processed via the routine {\em eis\_prep.pro} in the standard EIS software package with corrections for dark current, hot pixels, and cosmic ray hits. It observed the flaring region near the west limb for a period starting before 15:35 UT (flare onset) through 16:53 UT that covers the rise and early decay phases of the flare. The 2 arcsec slit of EIS was used to scan over an area of 240 arcsec $\times$ 304 arcsec from west to east with a course step of 3 arcsec, yielding a spatial resolution of 3 arcsec $\times$ 1 arcsec. It took about 9 min in each run with an exposure time of 5 s at each step.

Here we used the Fe XXIV 192.03 \AA~line with a formation temperature of 18 MK, in which the current sheet is most clearly visible. The Fe XXIV 192.03 \AA~line is believed to be blended with the Fe XI 192.02 \AA~line ($\sim$1 MK), but the blending could be safely ignored in large flares that contain hot plasmas. This can be verified by checking the relative strength of another line Fe XII 192.39 \AA~($\sim$1 MK) in the same spectral window, which is clearly separated from the Fe XXIV 192.03 \AA. Theoretically, the Fe XII 192.39 \AA~line is stronger than the Fe XI 192.02 \AA~line. Therefore, when the emission at 192.02/192.03 \AA~is greater than that at Fe XII 192.39 \AA, it should be mostly from the hot Fe XXIV 192.03 \AA~line. We examine all of the line profiles around the current sheet region and conclude that the emission is mainly contributed by the Fe XXIV 192.03 \AA~line \citep[also see][]{warren17,liying18}. In addition, we note that the Fe XXIV 192.03 \AA~line is saturated in some regions (mostly in flare loops) but not in the current sheet region under study. So we just discard those saturated line profiles in our study.

\begin{figure*} %%%%%%%%%%%%%%%%
\vspace{0.0\textwidth}
\centerline{\includegraphics[width=0.6\textwidth,clip=]{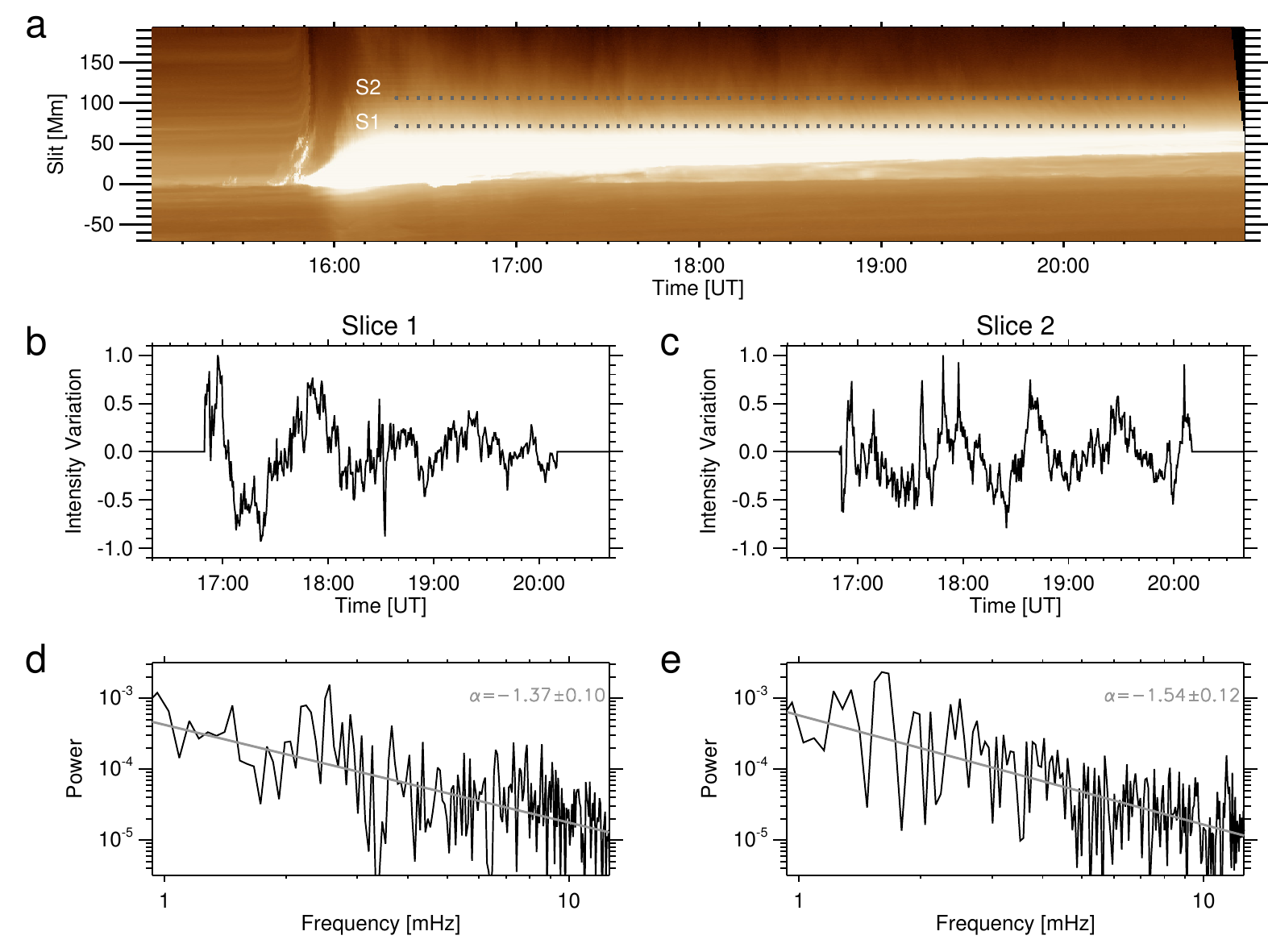}\hspace{0.0\textwidth} \vspace{-0.0\textwidth}}
\caption{(a) Time-distance plot of the AIA 193 {\AA} images along the slice 1 in Figure \ref{f3_193_131}. (b) and (c) The normalised 193 {\AA} intensity variations as a function of time at the two outflow regions indicated by S1 and S2 in panel a. It is also detrended with a moving average of 60 min in order to remove the feature of the decay reconnection process. (d) and (e) The power spectral densities of the intensities at S1 and S2 with the oblique lines indicating the power-law fitting to the range of 1--15 mHz.} \label{f3_193}
\end{figure*}

\begin{figure*} %%%%%%%%%%%%%%%%
\vspace{0.0\textwidth}
\centerline{\includegraphics[width=0.7\textwidth,clip=]{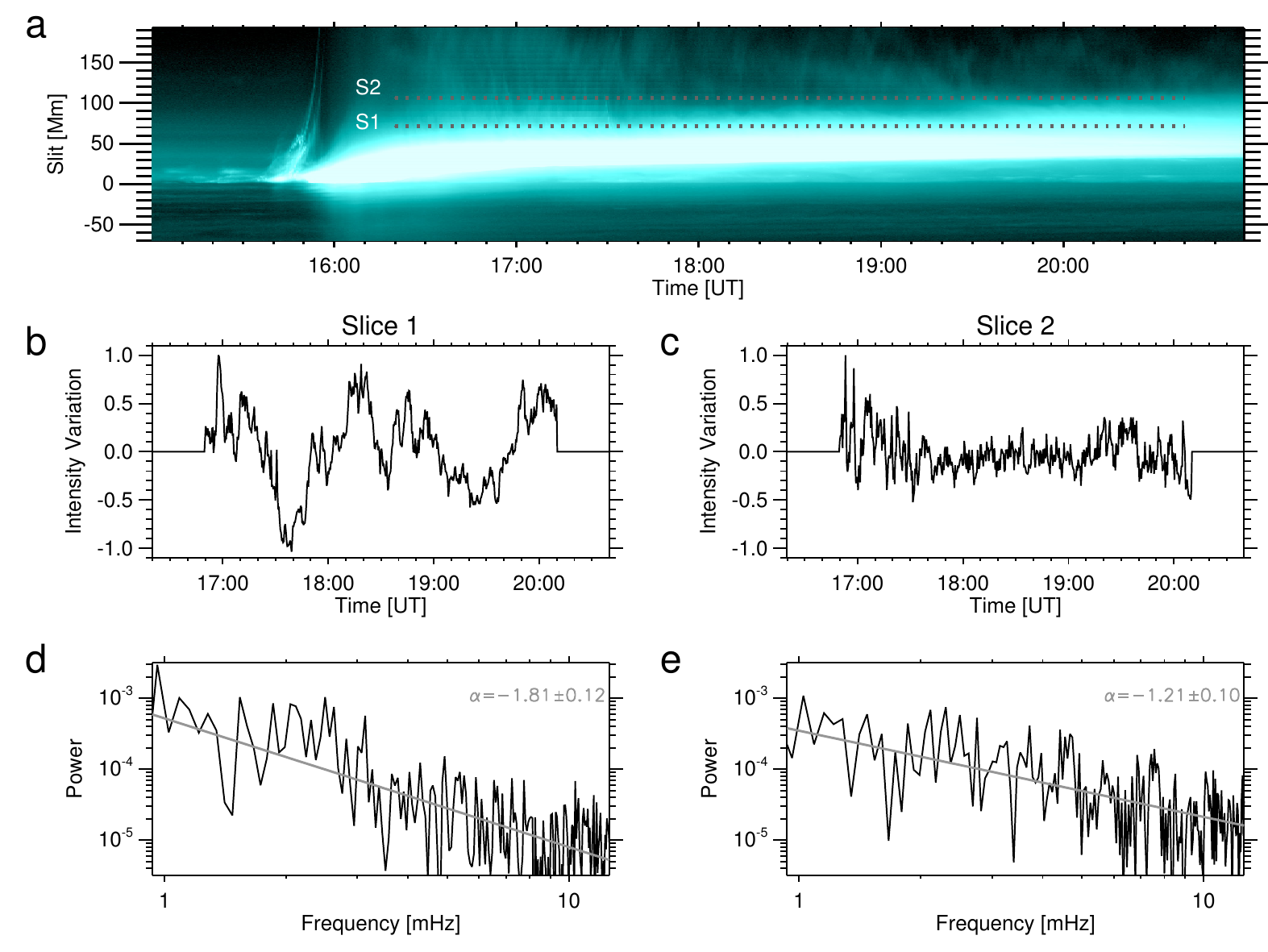}\hspace{0.0\textwidth} \vspace{-0.0\textwidth}}
\caption{Same as Figure \ref{f3_193} but for the AIA 131 {\AA} passband.} \label{f3_131}
\end{figure*}

The spectra of some other lines, for example, Fe XXIV 255.11 \AA~($\sim$18 MK), Fe XXIII 263.76 \AA~($\sim$14 MK), Fe XXII 253.17 \AA~($\sim$12 MK), Fe XVI 262.98 \AA~($\sim$3 MK), and Fe XV 284.16 \AA~($\sim$2 MK) are also presented (Figure \ref{fs_eis}). It is seen that the current sheet is only visible in the high temperature ($>$12 MK) lines, in particular in Fe XXIV 192.03 \AA, which is consistent with the AIA imaging observations. 

The Fe XXIV 192.03 \AA~line profiles show a good Gaussian shape, and we implement a single Gaussian fitting to obtain the non-thermal velocity by the formula 
\begin{equation}
  W=1.665\,\frac{\lambda}{c}\,{\sqrt{\frac{2\,k\,T_i}{M} + {\xi}^2}},
\end{equation}
where $W$ is the full width at half maximum of the spectral line, $\lambda$ is the line wavelength, $c$ is the speed of light, $k$ is the Boltzmann constant, $T_i$ is the ion temperature, and $M$ is the ion mass. The instrumental width of EIS (2.5 pixels, or 0.056 {\AA}) is also subtracted. Here we adopt a fixed thermal temperature of $T_i = T_{max} = 18$ MK. For comparison, we also use a DEM-weighted average temperature which is place-dependent to derive $\xi$ and find that the results are quite similar. The values are also consistent with that independently derived by \cite{warren17}. Please see \cite{liying18} for some selected Fe XXIV 192.03 \AA~line profiles and resulting fitting results.

\begin{figure*} %%%%%%%%%%%%%%%%
\vspace{0.0\textwidth}
\centerline{\includegraphics[width=0.7\textwidth,clip=]{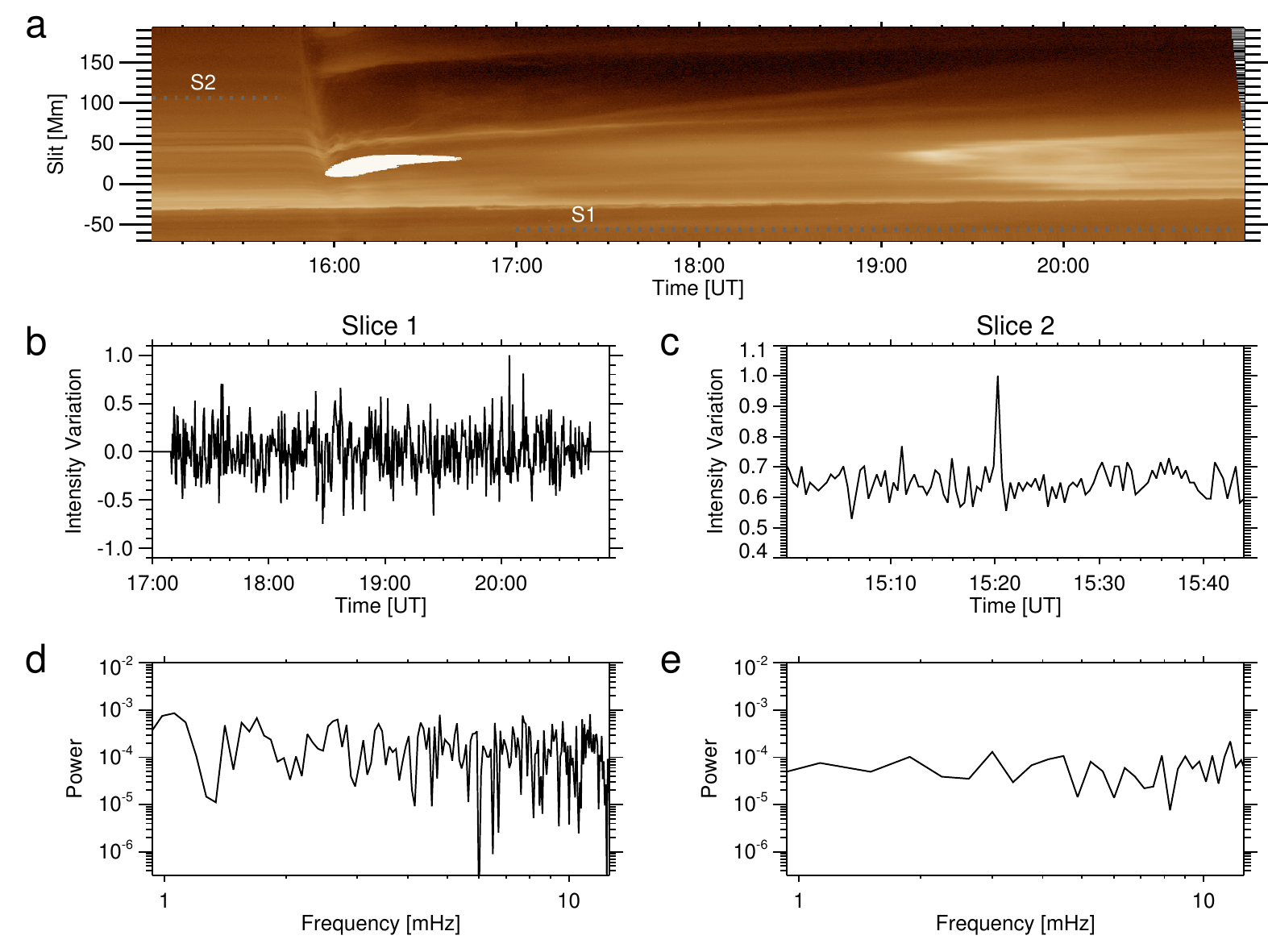}\hspace{0.0\textwidth} \vspace{-0.0\textwidth}}
\caption{(a) Time-distance plot of the AIA 193 {\AA} images along the slice 2 in Figure \ref{f3_193_131}. (b) and (c) The normalised 193 {\AA} intensity variations as a function of time at the two quiescent regions indicated by S1 and S2 in panel a. (d) and (e) The power spectral densities of the intensities at S1 and S2.} \label{f3_193_1}
\end{figure*}

\begin{figure*} %%%%%%%%%%%%%%%%
\vspace{0.0\textwidth}
\centerline{\includegraphics[width=0.7\textwidth,clip=]{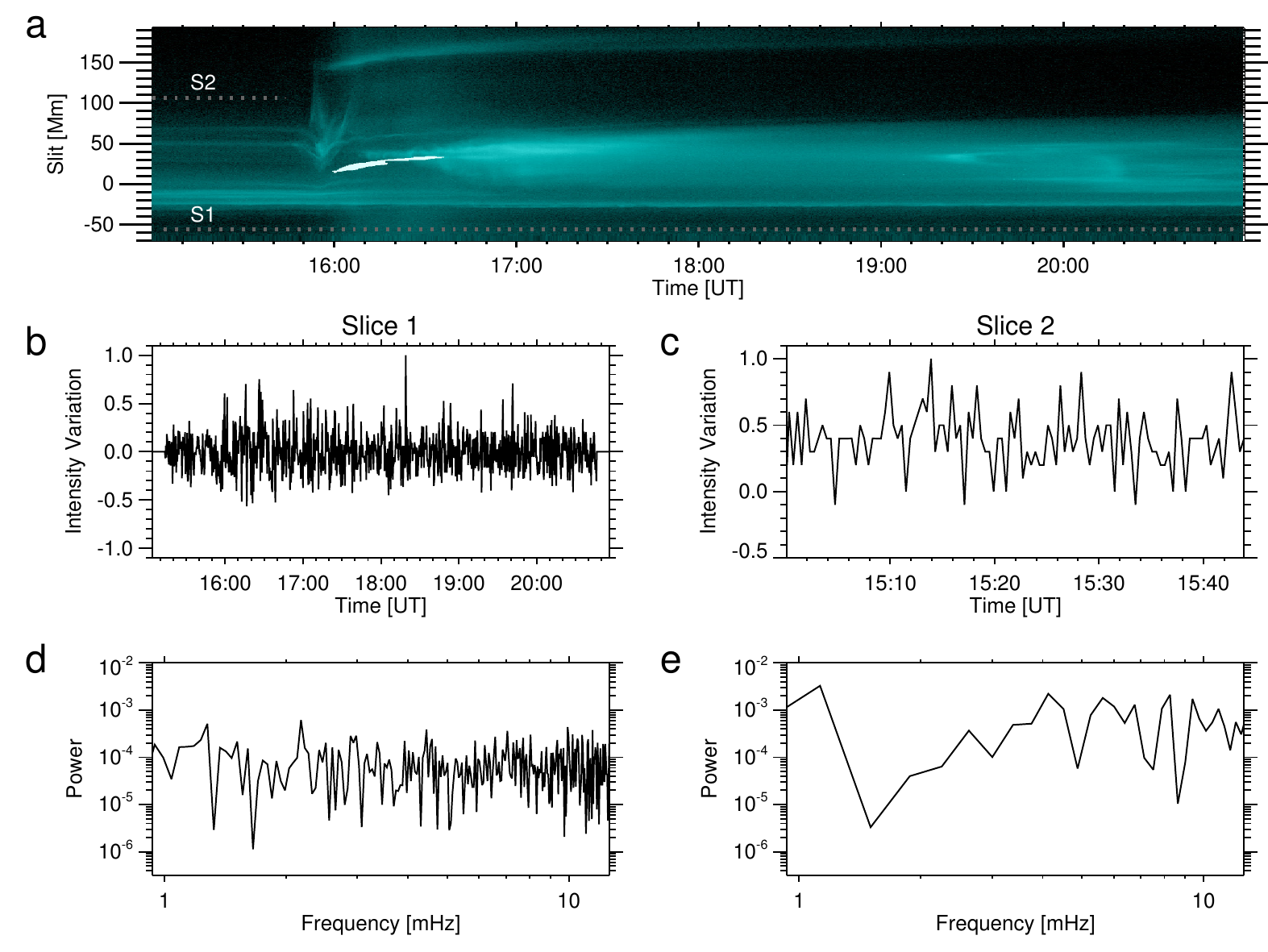}\hspace{0.0\textwidth} \vspace{-0.0\textwidth}}
\caption{Same as Figure \ref{f3_193_1} but for the AIA 131 {\AA} passband.} \label{f3_131_1}
\end{figure*}

\textbf{Fragmentation of the current sheet:}
The fragmentation of the current sheet is also examined at the different passbands and different times, as shown in Figure \ref{f_spectral_193} and \ref{f_spectral_131}. It can be seen that the spatial variation of the 193 {\AA} and 131 {\AA} intensity (along the current sheet) presents a power-law behaviour in spatial frequency domain (0.1--1.0 Mm$^{-1}$). The spectral index varies from --1.0 to --1.4. This type of fluctuation is also known as red ``noise", which is an intrinsic property of a random physical process, most likely due to turbulence, that can be described by a power-law spectrum with a negative slope \citep{vaughan05,inglis15,ning17}. It indicates that the current sheet has been fragmented into different scaled structures, most likely correspond to magnetic islands of different sizes. As shown in Figure \ref{f_spectral_193}b and \ref{f_spectral_131}b, the AIA 193 {\AA} and 131 {\AA} intensity has been detrended with a moving average of 10 Mm in order to remove the feature of the intensity decrease as away from the flare region. We also test the different moving average values (5--20 Mm) and find that derived spectral index is not seriously influenced. It is also worthy of noticing that, for the detrended data, the power in the low spatial frequency (e.g., $<$0.1 Mm$^{-1}$) can be artificially suppressed \citep{gruber11}, but which is not used here. Of course, as mentioned above, the AIA 193 {\AA} and 131 {\AA} intensity also include an uncertainty that mainly caused by non-linear effects of the AIA response function. The uncertainty may have somewhat effect on the spectral index \citep{ireland15}.

\begin{figure*}
\center {\includegraphics[width=15cm]{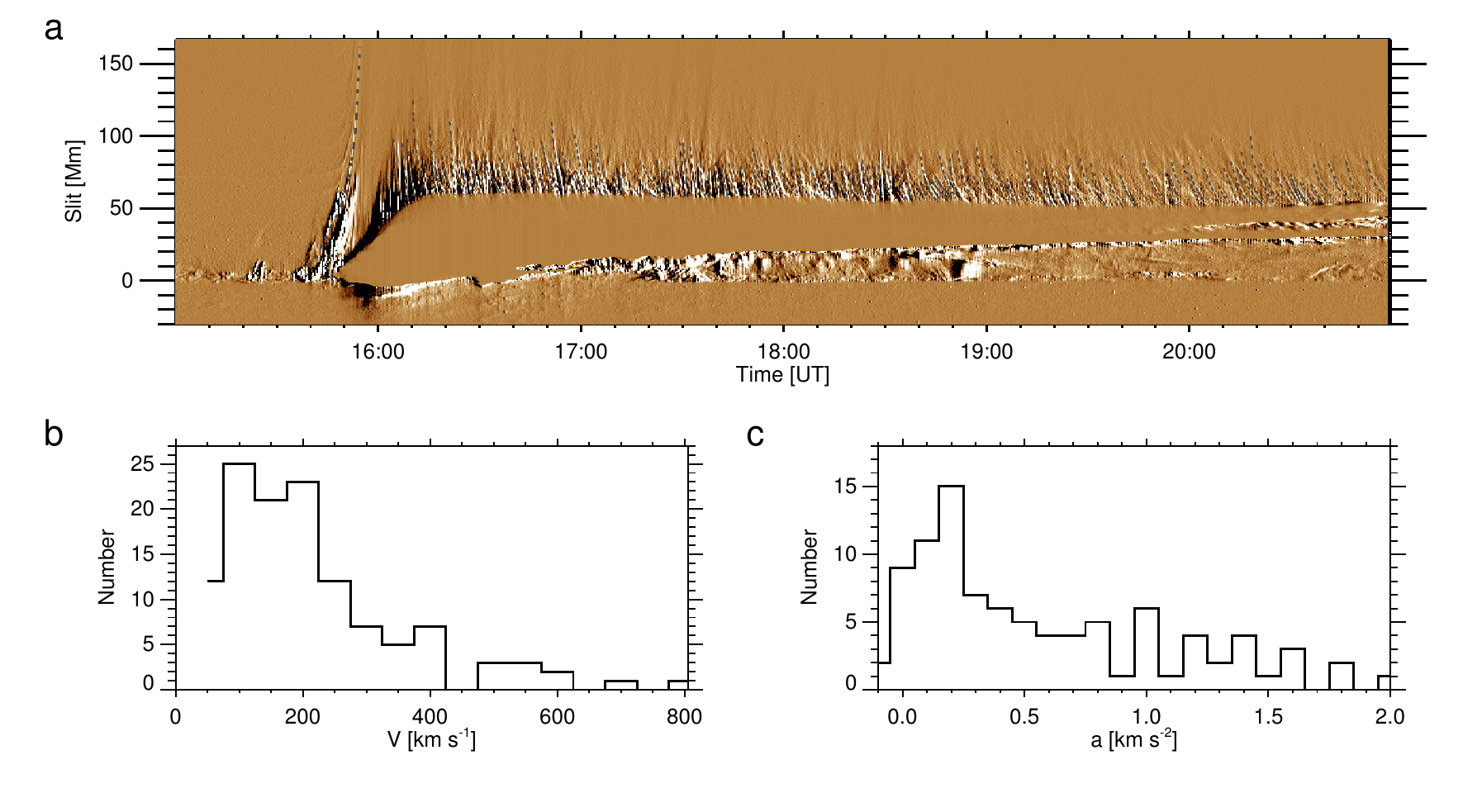}\hspace{-0.0\textwidth} \vspace{-0.03\textwidth}}
%\center {\includegraphics[width=15cm]{f4.pdf}\vspace{0.0\textwidth}}
\caption{(a) The time-distance plot of the AIA 193 \AA~running difference images with tracked trajectories of the erupting bubble (the first dashed line) and sunward moving outflow jets. (b and c) Histogram distributions of the initial velocities and the accelerations of the outflow jets.} \label{fs_track}
% (b) The velocity evolution for the erupting bubble.
\end{figure*}

\begin{figure*} %%%%%%%%%%%%%%%%
\vspace{0.0\textwidth}
\centerline{\includegraphics[width=0.68\textwidth,clip=]{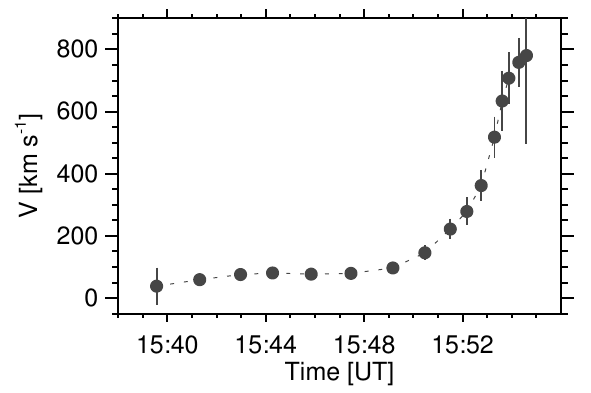}\hspace{0.0\textwidth} \vspace{-0.0\textwidth}}
\caption{The temporal evolution of the velocity of the erupting bubble with the errors indicated by the vertical bars.} \label{fs_vcme}
\end{figure*}

\begin{figure*} %%%%%%%%%%%%%%%%
\vspace{0.0\textwidth}
\centerline{\includegraphics[width=0.55\textwidth,clip=]{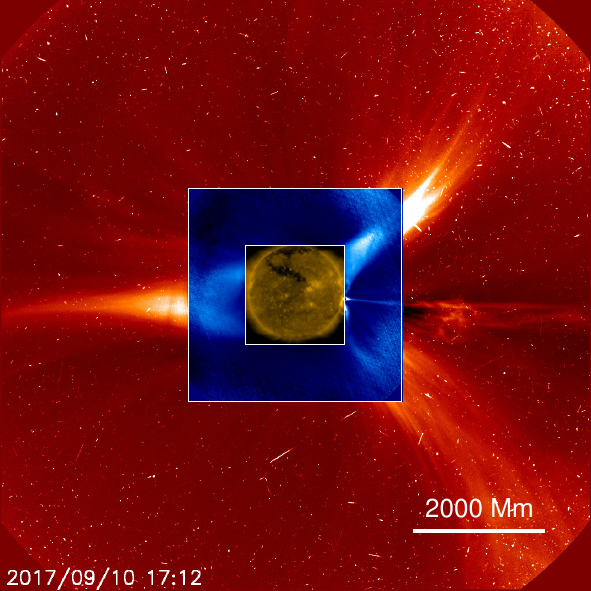}\hspace{0.0\textwidth} \vspace{-0.0\textwidth}}
\caption{A composition of the AIA 193 {\AA}, white-light K, and LASCO C2 images showing a largely extended current sheet at 17:12 UT.} \label{fs_c2}
\end{figure*}

\textbf{Intermittency and velocity diversity of the outflow jets:}
We also inspect the power spectrum of the temporal variations of the AIA 193 {\AA} and 131 {\AA} intensity at many other locations. Figure \ref{f3_193_131} shows two slices that we used for creating the time-distance plots. We find that the temporal variations of the intensity at almost all locations do display a power law distribution with spectral index distributing in the range of --1.2 to --1.8 (e.g., Figure \ref{f3_193} and Figure \ref{f3_131}). By contrast, for the quiescent regions and pre-flare regions, the spectrum is flat, which denotes white noise that is nearly frequency-independent and mainly originates from random signals (e.g., Figure \ref{f3_193_1} and Figure \ref{f3_131_1}). Similar to the spatial frequency analysis, the non-linear effects of the AIA passbands also influence the spectral index in the temporal frequency analysis \citep{ireland15}. Note that, the cadence of the AIA data is not exactly uniform, but which is found to be smaller than 0.05\%. After a carefully testing, we find that whether the non-uniformity is corrected or not does not significantly influence the FFT results. 

Using the time-distance plot of the AIA 193 {\AA} running difference images along the direction of the current sheet, we identified manually the trajectories of the sunward outflow jets, as shown in Figure \ref{fs_track}a. The initial speed is derived as an average of the first three points of the measured outflow speeds. The histogram distributions of the initial velocities and accelerations of the jets are displayed in Figure \ref{fs_track}b and \ref{fs_track}c. One can clearly see that both of them have a wide distribution.

\textbf{Height of the lower end and X-point of the current sheet:}
The heights of the CME bubble are measured in the AIA field of view. Applying the first order numerical derivative, the velocity as a function of time is derived (Figure \ref{fs_vcme}). One can see that the CME bubble experiences a slow rise phase of $\sim$10 min with an average speed of $\sim$70 km s$^{-1}$ and a fast acceleration phase with an acceleration of $\sim$2200 m s$^{-2}$ in the AIA field of view. The CME finally reaches a speed of over 3000 km s$^{-1}$ when leaving the LASCO field of view\footnote{https://cdaw.gsfc.nasa.gov/CME\_list/halo/halo.html}.

The lower end of the current sheet is estimated to be at the height of $\sim$100 Mm above the solar surface ($\sim$1100 arcsecs) at 16:15 UT, where the outflow jets are expelled. One hour later (17:15 UT), the lower end is also seen by the white-light K coronagraph. The height is determined to be $\sim$140 Mm. Considering that the lower end of the current sheet has an ascending motion as the CME erupts, its height will increase by 54 Mm in one hour if assuming a velocity of 15 km s$^{-1}$ \citep{mei12}. It roughly agrees with the difference between the heights derived in the EUV and white-light passbands at different instants. 

The theoretical model of flux-rope-induced CME/flare \citep{linjun00,mei12} also predicts an X-shaped null point existing in the current sheet. Magnetic islands are expected to run away from the null point along two opposite directions, manifesting as sunward outflow jets and anti-sunward fast moving blobs \citep[e.g.,][]{songhq12,chae17}, respectively. The jets and blobs have initial heights of $\sim$100 Mm and $\sim$180 Mm, respectively. It indicates that the height of the X-shaped null point is in the range of 100--180 Mm.

\textbf{Uncertainty in the reconnection rate:}
The reconnection rate is calculated as the ratio of the inflow velocity to the outflow velocity. We consider that the initial speed of the outflow jets is equivalent to the outflow speed. The error of the initial speed is about 100 km s$^{-1}$. Thus, the uncertainty in the reconnection rate can be up to 50\% when considering the initial velocities of 100 to 800 km s$^{-1}$ in the time period of 16:00--16:30 UT. If taking an average outflow speed of $\sim$300 km s$^{-1}$ and the inflow speed of 20 km s$^{-1}$, the average reconnection rate is about 0.07$\pm$0.03.

\textbf{Uncertainty in the length of the current sheet:}
We estimate the length of the current sheet based on the distributions of the brightness along the direction perpendicular to the current sheet (Figure \ref{f4}a). Figure \ref{f4}b shows that the brightness distributions at three slices have a similar profile with an FWHM being about 25 Mm. It is found that the FWHM is nearly uniform in between the two slices located at 840 Mm and 1240 Mm, respectively. Outside this region, the FWHM gets larger. Thus, the distance between the two slices is regarded as the length of the current sheet. Note that such a length is a lower limit. On the other hand, the LASCO observations show that the current sheet may even extend to a height of 8 $R_\odot$ at 17:12 UT, i.e., the edge of the C2 field of view where the blobs are still seen to move along the stretched bright structure by the erupting CME (Figure \ref{fs_c2}). It corresponds to a length of 4900 Mm. Such a length can be regarded an upper limit for the length of the current sheet. If the width remains to be 25 Mm, the upper limit of the length-to-width ratio for the current sheet is $\sim$200. Whatever the case may be, the length-to width-ratio is much larger than the theoretical threshold for tearing mode instability \citep[2$\pi$;][]{samtaney09}.


\begin{thebibliography}{83}
\expandafter\ifx\csname natexlab\endcsname\relax\def\natexlab#1{#1}\fi

\bibitem[{{Asai} {et~al.}(2004){Asai}, {Yokoyama}, {Shimojo}, \&
  {Shibata}}]{asai04}
{Asai}, A., {Yokoyama}, T., {Shimojo}, M., \& {Shibata}, K. 2004, \apjl, 605,
  L77

\bibitem[{{Badnell} {et~al.}(2003){Badnell}, {O'Mullane}, {Summers}, {Altun},
  {Bautista}, {Colgan}, {Gorczyca}, {Mitnik}, {Pindzola}, \&
  {Zatsarinny}}]{badnell03}
{Badnell}, N.~R., {et~al.} 2003, \aap, 406, 1151

\bibitem[{{Balbus} \& {Hawley}(1998)}]{balbus98}
{Balbus}, S.~A., \& {Hawley}, J.~F. 1998, Reviews of Modern Physics, 70, 1

\bibitem[{{B{\'a}rta} {et~al.}(2011){B{\'a}rta}, {B{\"u}chner}, {Karlick{\'y}},
  \& {Sk{\'a}la}}]{barta11}
{B{\'a}rta}, M., {B{\"u}chner}, J., {Karlick{\'y}}, M., \& {Sk{\'a}la}, J.
  2011, \apj, 737, 24

\bibitem[{{Bloom} {et~al.}(2011){Bloom}, {Giannios}, {Metzger}, {Cenko},
  {Perley}, {Butler}, {Tanvir}, {Levan}, {O'Brien}, {Strubbe}, {De Colle},
  {Ramirez-Ruiz}, {Lee}, {Nayakshin}, {Quataert}, {King}, {Cucchiara},
  {Guillochon}, {Bower}, {Fruchter}, {Morgan}, \& {van der Horst}}]{bloom11}
{Bloom}, J.~S., {et~al.} 2011, Science, 333, 203

\bibitem[{{Brueckner} {et~al.}(1995){Brueckner}, {Howard}, {Koomen},
  {Korendyke}, {et~al.}}]{brueckner95}
{Brueckner}, G.~E., {Howard}, R.~A., {Koomen}, M.~J., {Korendyke}, C.~M.,
  {et~al.} 1995, \solphys, 162, 357

\bibitem[{{Chae} {et~al.}(2017){Chae}, {Cho}, {Kwon}, \& {Lim}}]{chae17}
{Chae}, J., {Cho}, K., {Kwon}, R.-Y., \& {Lim}, E.-K. 2017, \apj, 841, 49

\bibitem[{{Chen}(2011)}]{chen11_review}
{Chen}, P.~F. 2011, Living Reviews in Solar Physics, 8, 1

\bibitem[{{Cheng} {et~al.}(2011){Cheng}, {Zhang}, {Liu}, \&
  {Ding}}]{cheng11_fluxrope}
{Cheng}, X., {Zhang}, J., {Liu}, Y., \& {Ding}, M.~D. 2011, \apjl, 732, L25

\bibitem[{{Cheng} {et~al.}(2012){Cheng}, {Zhang}, {Saar}, \&
  {Ding}}]{cheng12_dem}
{Cheng}, X., {Zhang}, J., {Saar}, S.~H., \& {Ding}, M.~D. 2012, \apj, 761, 62

\bibitem[{{Cheung} {et~al.}(2015){Cheung}, {Boerner}, {Schrijver}, {Testa},
  {Chen}, {Peter}, \& {Malanushenko}}]{cheung15_dem}
{Cheung}, M.~C.~M., {Boerner}, P., {Schrijver}, C.~J., {Testa}, P., {Chen}, F.,
  {Peter}, H., \& {Malanushenko}, A. 2015, \apj, 807, 143

\bibitem[{{Ciaravella} \& {Raymond}(2008)}]{ciaravella08}
{Ciaravella}, A., \& {Raymond}, J.~C. 2008, \apj, 686, 1372

\bibitem[{{Ciaravella} {et~al.}(2003){Ciaravella}, {Raymond}, {van
  Ballegooijen}, {Strachan}, {Vourlidas}, {Li}, {Chen}, \&
  {Panasyuk}}]{ciaravella03}
{Ciaravella}, A., {Raymond}, J.~C., {van Ballegooijen}, A., {Strachan}, L.,
  {Vourlidas}, A., {Li}, J., {Chen}, J., \& {Panasyuk}, A. 2003, \apj, 597,
  1118

\bibitem[{{Culhane} {et~al.}(2007){Culhane}, {Harra}, {James}, {Al-Janabi},
  {Bradley}, {Chaudry}, {Rees}, {Tandy}, {Thomas}, {Whillock}, {Winter},
  {Doschek}, {Korendyke}, {Brown}, {Myers}, {Mariska}, {Seely}, {Lang}, {Kent},
  {Shaughnessy}, {Young}, {Simnett}, {Castelli}, {Mahmoud}, {Mapson-Menard},
  {Probyn}, {Thomas}, {Davila}, {Dere}, {Windt}, {Shea}, {Hagood}, {Moye},
  {Hara}, {Watanabe}, {Matsuzaki}, {Kosugi}, {Hansteen}, \&
  {Wikstol}}]{culhane07}
{Culhane}, J.~L., {et~al.} 2007, \solphys, 243, 19

\bibitem[{{Doschek} {et~al.}(2014){Doschek}, {McKenzie}, \&
  {Warren}}]{Doschek14}
{Doschek}, G.~A., {McKenzie}, D.~E., \& {Warren}, H.~P. 2014, \apj, 788, 26

\bibitem[{{Forbes} \& {Acton}(1996)}]{forbes96}
{Forbes}, T.~G., \& {Acton}, L.~W. 1996, \apj, 459, 330

\bibitem[{{Freeland} \& {Handy}(1998)}]{Freeland98}
{Freeland}, S.~L., \& {Handy}, B.~N. 1998, \solphys, 182, 497

\bibitem[{{Furth} {et~al.}(1963){Furth}, {Killeen}, \& {Rosenbluth}}]{furth63}
{Furth}, H.~P., {Killeen}, J., \& {Rosenbluth}, M.~N. 1963, Physics of Fluids,
  6, 459

\bibitem[{{Golub} {et~al.}(2007){Golub}, {Deluca}, {Austin}, {Bookbinder},
  {et~al.}}]{golub07}
{Golub}, L., {Deluca}, E., {Austin}, G., {Bookbinder}, J., {et~al.} 2007,
  \solphys, 243, 63

\bibitem[{{Gruber} {et~al.}(2011){Gruber}, {Lachowicz}, {Bissaldi}, {Briggs},
  {Connaughton}, {Greiner}, {van der Horst}, {Kanbach}, {Rau}, {Bhat}, {Diehl},
  {von Kienlin}, {Kippen}, {Meegan}, {Paciesas}, {Preece}, \&
  {Wilson-Hodge}}]{gruber11}
{Gruber}, D., {et~al.} 2011, \aap, 533, A61

\bibitem[{{Guo} {et~al.}(2014){Guo}, {Huang}, {Bhattacharjee}, \&
  {Innes}}]{guolj14}
{Guo}, L.-J., {Huang}, Y.-M., {Bhattacharjee}, A., \& {Innes}, D.~E. 2014,
  \apjl, 796, L29

\bibitem[{{Hannah} \& {Kontar}(2012)}]{hannah12}
{Hannah}, I.~G., \& {Kontar}, E.~P. 2012, \aap, 539, A146

\bibitem[{{Hanneman} \& {Reeves}(2014)}]{Hanneman14}
{Hanneman}, W.~J., \& {Reeves}, K.~K. 2014, \apj, 786, 95

\bibitem[{{Imada} {et~al.}(2011){Imada}, {Murakami}, {Watanabe}, {Hara}, \&
  {Shimizu}}]{imada11}
{Imada}, S., {Murakami}, I., {Watanabe}, T., {Hara}, H., \& {Shimizu}, T. 2011,
  \apj, 742, 70

\bibitem[{{Inglis} {et~al.}(2015){Inglis}, {Ireland}, \&
  {Dominique}}]{inglis15}
{Inglis}, A.~R., {Ireland}, J., \& {Dominique}, M. 2015, \apj, 798, 108

\bibitem[{{Innes} {et~al.}(2015){Innes}, {Guo}, {Huang}, \&
  {Bhattacharjee}}]{innes15}
{Innes}, D.~E., {Guo}, L.-J., {Huang}, Y.-M., \& {Bhattacharjee}, A. 2015,
  \apj, 813, 86

\bibitem[{{Ireland} {et~al.}(2015){Ireland}, {McAteer}, \&
  {Inglis}}]{ireland15}
{Ireland}, J., {McAteer}, R.~T.~J., \& {Inglis}, A.~R. 2015, \apj, 798, 1

\bibitem[{{Judge}(2010)}]{judge10}
{Judge}, P.~G. 2010, \apj, 708, 1238

\bibitem[{{Kontar} {et~al.}(2017){Kontar}, {Perez}, {Harra}, {Kuznetsov},
  {Emslie}, {Jeffrey}, {Bian}, \& {Dennis}}]{kontar17}
{Kontar}, E.~P., {Perez}, J.~E., {Harra}, L.~K., {Kuznetsov}, A.~A., {Emslie},
  A.~G., {Jeffrey}, N.~L.~S., {Bian}, N.~H., \& {Dennis}, B.~R. 2017, Physical
  Review Letters, 118, 155101

\bibitem[{{Kosugi} {et~al.}(2007){Kosugi}, {Matsuzaki}, {Sakao}, {Shimizu},
  {Sone}, {Tachikawa}, {Hashimoto}, {Minesugi}, {Ohnishi}, {Yamada}, {Tsuneta},
  {Hara}, {Ichimoto}, {Suematsu}, {Shimojo}, {Watanabe}, {Shimada}, {Davis},
  {Hill}, {Owens}, {Title}, {Culhane}, {Harra}, {Doschek}, \&
  {Golub}}]{kosugi07}
{Kosugi}, T., {et~al.} 2007, \solphys, 243, 3

\bibitem[{{Kowal} {et~al.}(2009){Kowal}, {Lazarian}, {Vishniac}, \&
  {Otmianowska-Mazur}}]{kowal09}
{Kowal}, G., {Lazarian}, A., {Vishniac}, E.~T., \& {Otmianowska-Mazur}, K.
  2009, \apj, 700, 63

\bibitem[{{Lazarian} \& {Vishniac}(1999)}]{lazarian99}
{Lazarian}, A., \& {Vishniac}, E.~T. 1999, \apj, 517, 700

\bibitem[{{Lemen} {et~al.}(2012){Lemen}, {Title}, {Akin}, {Boerner}, {Chou},
  {Drake}, {Duncan}, {Edwards}, {Friedlaender}, {Heyman}, {Hurlburt}, {Katz},
  {Kushner}, {Levay}, {Lindgren}, {Mathur}, {McFeaters}, {Mitchell}, {Rehse},
  {Schrijver}, {Springer}, {Stern}, {Tarbell}, {Wuelser}, {Wolfson}, {Yanari},
  {Bookbinder}, {Cheimets}, {Caldwell}, {Deluca}, {Gates}, {Golub}, {Park},
  {Podgorski}, {Bush}, {Scherrer}, {Gummin}, {Smith}, {Auker}, {Jerram},
  {Pool}, {Soufli}, {Windt}, {Beardsley}, {Clapp}, {Lang}, \&
  {Waltham}}]{lemen12}
{Lemen}, J.~R., {et~al.} 2012, \solphys, 275, 17

\bibitem[{{Li} {et~al.}(2016){Li}, {Zhang}, {Peter}, {Priest}, {Chen}, {Guo},
  {Chen}, \& {Mackay}}]{lileping16_np}
{Li}, L., {Zhang}, J., {Peter}, H., {Priest}, E., {Chen}, H., {Guo}, L.,
  {Chen}, F., \& {Mackay}, D. 2016, Nature Physics, 12, 847

\bibitem[{{Li} {et~al.}(2017){Li}, {Sun}, {Ding}, {Qiu}, \&
  {Priest}}]{liying17}
{Li}, Y., {Sun}, X., {Ding}, M.~D., {Qiu}, J., \& {Priest}, E.~R. 2017, \apj,
  835, 190

\bibitem[{{Li} {et~al.}(2018){Li}, {Xue}, {Ding}, {Cheng}, {Su}, {Feng},
  {Hong}, {Li}, \& {Gan}}]{liying18}
{Li}, Y., {et~al.} 2018, \apjl, 853, L15

\bibitem[{{Lin} \& {Forbes}(2000)}]{linjun00}
{Lin}, J., \& {Forbes}, T.~G. 2000, \jgr, 105, 2375

\bibitem[{{Lin} {et~al.}(2005){Lin}, {Ko}, {Sui}, {Raymond}, {Stenborg},
  {Jiang}, {Zhao}, \& {Mancuso}}]{linjun05}
{Lin}, J., {Ko}, Y.-K., {Sui}, L., {Raymond}, J.~C., {Stenborg}, G.~A.,
  {Jiang}, Y., {Zhao}, S., \& {Mancuso}, S. 2005, \apj, 622, 1251

\bibitem[{{Lin} {et~al.}(2009){Lin}, {Li}, {Ko}, \& {Raymond}}]{linj09}
{Lin}, J., {Li}, J., {Ko}, Y.-K., \& {Raymond}, J.~C. 2009, \apj, 693, 1666

\bibitem[{{Lin} {et~al.}(2015){Lin}, {Murphy}, {Shen}, {Raymond}, {Reeves},
  {Zhong}, {Wu}, \& {Li}}]{linj15}
{Lin}, J., {Murphy}, N.~A., {Shen}, C., {Raymond}, J.~C., {Reeves}, K.~K.,
  {Zhong}, J., {Wu}, N., \& {Li}, Y. 2015, \ssr, 194, 237

\bibitem[{{Ling} {et~al.}(2014){Ling}, {Webb}, {Burkepile}, \&
  {Cliver}}]{ling14}
{Ling}, A.~G., {Webb}, D.~F., {Burkepile}, J.~T., \& {Cliver}, E.~W. 2014,
  \apj, 784, 91

\bibitem[{{Liu}(2013)}]{liurui13}
{Liu}, R. 2013, \mnras, 434, 1309

\bibitem[{{Liu} {et~al.}(2013){Liu}, {Chen}, \& {Petrosian}}]{liuw13}
{Liu}, W., {Chen}, Q., \& {Petrosian}, V. 2013, \apj, 767, 168

\bibitem[{{Liu} {et~al.}(2018){Liu}, {Jin}, {Downs}, {Ofman}, {Cheung}, \&
  {Nitta}}]{liuwei18}
{Liu}, W., {Jin}, M., {Downs}, C., {Ofman}, L., {Cheung}, M.~C.~M., \& {Nitta},
  N.~V. 2018, ArXiv e-prints

\bibitem[{{Masuda} {et~al.}(1994){Masuda}, {Kosugi}, {Hara}, {Tsuneta}, \&
  {Ogawara}}]{masuda94}
{Masuda}, S., {Kosugi}, T., {Hara}, H., {Tsuneta}, S., \& {Ogawara}, Y. 1994,
  \nat, 371, 495

\bibitem[{{McKenzie}(2000)}]{mckenzie02}
{McKenzie}, D.~E. 2000, \solphys, 195, 381

\bibitem[{{McKenzie}(2013)}]{mckenzie13}
---. 2013, \apj, 766, 39

\bibitem[{{Mei} {et~al.}(2012){Mei}, {Shen}, {Wu}, {Lin}, {Murphy}, \&
  {Roussev}}]{mei12}
{Mei}, Z., {Shen}, C., {Wu}, N., {Lin}, J., {Murphy}, N.~A., \& {Roussev},
  I.~I. 2012, \mnras, 425, 2824

\bibitem[{{Ning}(2017)}]{ning17}
{Ning}, Z. 2017, \solphys, 292, 11

\bibitem[{{Nishizuka} {et~al.}(2009){Nishizuka}, {Asai}, {Takasaki},
  {Kurokawa}, \& {Shibata}}]{nishizuka09}
{Nishizuka}, N., {Asai}, A., {Takasaki}, H., {Kurokawa}, H., \& {Shibata}, K.
  2009, \apjl, 694, L74

\bibitem[{{Pesnell} {et~al.}(2012){Pesnell}, {Thompson}, \&
  {Chamberlin}}]{pesnell12}
{Pesnell}, W.~D., {Thompson}, B.~J., \& {Chamberlin}, P.~C. 2012, \solphys,
  275, 3

\bibitem[{{Petschek}(1964)}]{petschek64}
{Petschek}, H.~E. 1964, NASA Special Publication, 50, 425

\bibitem[{{Phan} {et~al.}(2006){Phan}, {Gosling}, {Davis}, {Skoug},
  {{\O}ieroset}, {Lin}, {Lepping}, {McComas}, {Smith}, {Reme}, \&
  {Balogh}}]{phan06}
{Phan}, T.~D., {et~al.} 2006, \nat, 439, 175

\bibitem[{{Priest}(2014)}]{priest14}
{Priest}, E. 2014, {Magnetohydrodynamics of the Sun}

\bibitem[{{Priest} \& {Forbes}(2002)}]{priest02}
{Priest}, E.~R., \& {Forbes}, T.~G. 2002, \aapr, 10, 313

\bibitem[{{Reeves} {et~al.}(2015){Reeves}, {McCauley}, \& {Tian}}]{reeves15}
{Reeves}, K.~K., {McCauley}, P.~I., \& {Tian}, H. 2015, \apj, 807, 7

\bibitem[{{Samtaney} {et~al.}(2009){Samtaney}, {Loureiro}, {Uzdensky},
  {Schekochihin}, \& {Cowley}}]{samtaney09}
{Samtaney}, R., {Loureiro}, N.~F., {Uzdensky}, D.~A., {Schekochihin}, A.~A., \&
  {Cowley}, S.~C. 2009, Physical Review Letters, 103, 105004

\bibitem[{{Savage} \& {McKenzie}(2011)}]{savage11}
{Savage}, S.~L., \& {McKenzie}, D.~E. 2011, \apj, 730, 98

\bibitem[{{Savage} {et~al.}(2010){Savage}, {McKenzie}, {Reeves}, {Forbes}, \&
  {Longcope}}]{savage10}
{Savage}, S.~L., {McKenzie}, D.~E., {Reeves}, K.~K., {Forbes}, T.~G., \&
  {Longcope}, D.~W. 2010, \apj, 722, 329

\bibitem[{{Scott} {et~al.}(2016){Scott}, {McKenzie}, \& {Longcope}}]{scott16}
{Scott}, R.~B., {McKenzie}, D.~E., \& {Longcope}, D.~W. 2016, \apj, 819, 56

\bibitem[{{Seaton} {et~al.}(2017){Seaton}, {Bartz}, \& {Darnel}}]{seaton17}
{Seaton}, D.~B., {Bartz}, A.~E., \& {Darnel}, J.~M. 2017, \apj, 835, 139

\bibitem[{{Shen} {et~al.}(2011){Shen}, {Lin}, \& {Murphy}}]{shencc11}
{Shen}, C., {Lin}, J., \& {Murphy}, N.~A. 2011, \apj, 737, 14

\bibitem[{{Shibata} {et~al.}(1995){Shibata}, {Masuda}, {Shimojo}, {Hara},
  {Yokoyama}, {Tsuneta}, {Kosugi}, \& {Ogawara}}]{shibata95}
{Shibata}, K., {Masuda}, S., {Shimojo}, M., {Hara}, H., {Yokoyama}, T.,
  {Tsuneta}, S., {Kosugi}, T., \& {Ogawara}, Y. 1995, \apjl, 451, L83

\bibitem[{{Shibata} \& {Tanuma}(2001)}]{shibata01}
{Shibata}, K., \& {Tanuma}, S. 2001, Earth, Planets, and Space, 53, 473

\bibitem[{{Song} {et~al.}(2012){Song}, {Kong}, {Chen}, {Li}, {Li}, {Feng}, \&
  {Xia}}]{songhq12}
{Song}, H.~Q., {Kong}, X.~L., {Chen}, Y., {Li}, B., {Li}, G., {Feng}, S.~W., \&
  {Xia}, L.~D. 2012, \solphys, 276, 261

\bibitem[{{Strauss}(1988)}]{strauss88}
{Strauss}, H.~R. 1988, \apj, 326, 412

\bibitem[{{Sturrock}(1966)}]{sturrock66}
{Sturrock}, P.~A. 1966, \nat, 211, 695

\bibitem[{{Su} {et~al.}(2013){Su}, {Veronig}, {Holman}, {Dennis}, {Wang},
  {Temmer}, \& {Gan}}]{suyang13}
{Su}, Y., {Veronig}, A.~M., {Holman}, G.~D., {Dennis}, B.~R., {Wang}, T.,
  {Temmer}, M., \& {Gan}, W. 2013, Nature Physics, 9, 489

\bibitem[{{Sui} \& {Holman}(2003)}]{sui03}
{Sui}, L., \& {Holman}, G.~D. 2003, \apjl, 596, L251

\bibitem[{{Summers}(1974)}]{summers74}
{Summers}, H.~P. 1974, \mnras, 169, 663

\bibitem[{{Takasao} {et~al.}(2012){Takasao}, {Asai}, {Isobe}, \&
  {Shibata}}]{takasao12}
{Takasao}, S., {Asai}, A., {Isobe}, H., \& {Shibata}, K. 2012, \apjl, 745, L6

\bibitem[{{Takasao} {et~al.}(2016){Takasao}, {Asai}, {Isobe}, \&
  {Shibata}}]{takasao16}
---. 2016, \apj, 828, 103

\bibitem[{{Vaughan}(2005)}]{vaughan05}
{Vaughan}, S. 2005, \aap, 431, 391

\bibitem[{{Wang} {et~al.}(2017){Wang}, {Sim{\~o}es}, {Jeffrey}, {Fletcher},
  {Wright}, \& {Hannah}}]{wangjt17}
{Wang}, J., {Sim{\~o}es}, P.~J.~A., {Jeffrey}, N.~L.~S., {Fletcher}, L.,
  {Wright}, P.~J., \& {Hannah}, I.~G. 2017, \apjl, 847, L1

\bibitem[{{Warren} {et~al.}(2018){Warren}, {Brooks}, {Ugarte-Urra}, {Reep},
  {Crump}, \& {Doschek}}]{warren17}
{Warren}, H.~P., {Brooks}, D.~H., {Ugarte-Urra}, I., {Reep}, J.~W., {Crump},
  N.~A., \& {Doschek}, G.~A. 2018, \apj, 854, 122

\bibitem[{{Weber} {et~al.}(2004){Weber}, {Deluca}, {Golub}, \&
  {Sette}}]{weber04}
{Weber}, M.~A., {Deluca}, E.~E., {Golub}, L., \& {Sette}, A.~L. 2004, in IAU
  Symposium, Vol. 223, Multi-Wavelength Investigations of Solar Activity, ed.
  A.~V. {Stepanov}, E.~E. {Benevolenskaya}, \& A.~G. {Kosovichev}, 321--328

\bibitem[{{Xue} {et~al.}(2016){Xue}, {Yan}, {Cheng}, {Yang}, {Su}, {Kliem},
  {Zhang}, {Liu}, {Bi}, {Xiang}, {Yang}, \& {Zhao}}]{xuezhike16}
{Xue}, Z., {et~al.} 2016, Nature Communications, 7, 11837

\bibitem[{{Yamada} {et~al.}(2010){Yamada}, {Kulsrud}, \& {Ji}}]{yamada10}
{Yamada}, M., {Kulsrud}, R., \& {Ji}, H. 2010, Reviews of Modern Physics, 82,
  603

\bibitem[{{Yan} {et~al.}(2018){Yan}, {Yang}, {Xue}, {Mei}, {Kong}, {Wang}, \&
  {Li}}]{yanxl18_cs}
{Yan}, X.~L., {Yang}, L.~H., {Xue}, Z.~K., {Mei}, Z.~X., {Kong}, D.~F., {Wang},
  J.~C., \& {Li}, Q.~L. 2018, \apjl, 853, L18

\bibitem[{{Yang} {et~al.}(2015){Yang}, {Zhang}, \& {Xiang}}]{yangshuhong15}
{Yang}, S., {Zhang}, J., \& {Xiang}, Y. 2015, \apjl, 798, L11

\bibitem[{{Yokoyama} {et~al.}(2001){Yokoyama}, {Akita}, {Morimoto}, {Inoue}, \&
  {Newmark}}]{yokoyama01}
{Yokoyama}, T., {Akita}, K., {Morimoto}, T., {Inoue}, K., \& {Newmark}, J.
  2001, \apjl, 546, L69

\bibitem[{{Zhang} {et~al.}(2001){Zhang}, {Dere}, {Howard}, {Kundu}, \&
  {White}}]{zhang01}
{Zhang}, J., {Dere}, K.~P., {Howard}, R.~A., {Kundu}, M.~R., \& {White}, S.~M.
  2001, \apj, 559, 452

\bibitem[{{Zhu} {et~al.}(2016){Zhu}, {Liu}, {Alexander}, \&
  {McAteer}}]{zhucm16}
{Zhu}, C., {Liu}, R., {Alexander}, D., \& {McAteer}, R.~T.~J. 2016, \apjl, 821,
  L29

\end{thebibliography}
\end{document}